\documentclass[final,5p,times,twocolumn,12pt]{elsarticle}

\newif\ifBWfigs
\BWfigsfalse

\usepackage{amssymb}
\usepackage[symbol]{footmisc} 
\usepackage{lineno}

\usepackage[hidelinks]{hyperref}

\usepackage{booktabs}
\usepackage{adjustbox}
\usepackage{graphicx}
\usepackage{dcolumn}
\usepackage{bm}
\usepackage{url}
\usepackage{amsmath}
\usepackage{verbatim}  
\usepackage{relsize}
\usepackage{xspace}
\usepackage{fixltx2e} 
\usepackage[dvipsnames]{xcolor}
\usepackage{arydshln}

\newcommand{\pb}{$^{208}\mathrm{Pb}^{82+}$}

\newcommand{\nb}{nb$^{-1}$\xspace}

\newcommand{\Lu}{\mathcal{L}}
\newcommand{\Ltot}{\Lu_\mathrm{tot}}

\journal{Nuclear Instruments and Methods in Physics Research Section A}

\begin{document}

\begin{frontmatter}



\title{Charting the Luminosity Capabilities of the CERN Large Hadron Collider with Various Nuclear Species} 


\author[epfl,cern]{E. Waagaard}
\ead{elias.walter.waagaard@cern.ch}
\author[cern]{R. Bruce\corref{cor1}}
\ead{roderik.bruce@cern.ch}
\author[cern]{R. Alemany Fern\'andez}
\author[cern]{H. Bartosik}
\author[cern,gsi]{J.M. Jowett}
\author[cern]{N. Triantafyllou}
\cortext[cor1]{Corresponding author.}
\address[epfl]{EPFL, Rte Cantonale, 1015 Lausanne, Switzerland}
\address[cern]{CERN, 1211 Geneva 23, Switzerland}
\address[gsi]{GSI Helmholtzzentrum für Schwerionenforschung GmbH, Darmstadt, Germany}

\begin{abstract}
The Large Hadron Collider (LHC) at CERN 
has been instrumental in recent advances in experimental high energy physics by colliding beams of protons and heavier nuclei at unprecedented energies. The present heavy-ion programme is based mainly on colliding lead nuclei. For future ion runs, there is strong interest to achieve a significantly higher integrated nucleon-nucleon luminosity, which might be achieved through collisions of species other than Pb. In this paper, we explore the intensity and nucleon-nucleon luminosity projections in the LHC for a selection of  ion species ranging from He to Xe, and including Pb as reference. Alternative beam production schemes are investigated as a way to mitigate effects such as space charge that degrade the beam quality in the LHC injectors. In the most optimistic scenarios, we find up to about a factor~4 improvement in integrated nucleon-nucleon luminosity for a typical future one-month run, with respect to the present Pb programme. We also outline a future study programme and experiments to test the assumptions and refine the simulated projections put forward in this article. 
\end{abstract}


\end{frontmatter}


\section{Introduction}
\label{sec:intro}
The Large Hadron Collider (LHC)~\cite{lhcdesignV1} is a 27~km synchrotron, designed to collide 7~TeV protons, or nuclei (fully-stripped ions) of equal magnetic rigidity, at four interaction points inside experimental detectors: ATLAS, ALICE, CMS, and LHCb. The LHC has typically operated with \pb-ion beams for about one month per year during Run~1\footnote{Here we follow the usual CERN convention referring to annual operating periods as ``runs'' and multi-year operating periods between long shutdowns as (capitalised) ``Runs''.}   (2010--2013) and Run~2 (2015--2018). Two runs have also been carried out so far in Run~3 (which started in 2022). In total, there have been six one-month Pb-Pb 
runs~\cite{ipac11_jowett_fist_Pb_run,jowett16_ipac,jowett19_ipac,jowett19_evian}, resulting in a delivered integrated luminosity of about 5.6~\nb at ALICE and 6.3~\nb at ATLAS and CMS, out of which 3.7--3.9~\nb were collected in Run~3 alone due to the higher performance achieved from upgrades for the High-Luminosity (HL-LHC)~\cite{Alonso2020,bruce20_HL_ion_report, liu_TDR_ions}. The HL-LHC upgrades for heavy-ions are already available since the start of Run 3, while the proton upgrades are scheduled to become available for Run~4 (presently foreseen for 2030--2033).

As of today,  heavy-ion operation is foreseen to continue throughout Run 3 (ending mid-2026), Run 4 (foreseen for 2030--2033), and Run~5 (foreseen for 2036--2041) with yearly runs of several weeks of either Pb-Pb or p-Pb collisions. 
The target integrated luminosity is 13~\nb for Pb-Pb in Run~3 and Run~4 combined at each of the three experiments ALICE, ATLAS and CMS, as outlined in~\cite{YR_WG5_2018} by the Working Group 5 (WG5) on the Physics of the HL-LHC, 
assuming the machine scenario presented in~\cite{bruce20_HL_ion_report,bruce21_EPJplus_ionLumi}. Additional goals have been set for p-Pb operation before the end of Run~4~\cite{YR_WG5_2018,bruce21_EPJplus_ionLumi}.

An extended ion programme is being studied for Run~5~\cite{YR_WG5_2018}. A new detector for the ALICE experiment (ALICE 3) is under study to further investigate the properties of the quark-gluon plasma and of strongly interacting matter~\cite{alice_2022_letter_of_intent}. Extending ion operation using this new detector has the main aim of reaching a significantly higher integrated nucleon-nucleon luminosity than in all previous Runs. Numerous improvements of the injectors in recent years through the LHC Injector Upgrade (LIU) project~\cite{liu_TDR_ions} have increased the ion beam intensity available for the LHC, by almost doubling the number of Pb bunches in the LHC and increasing the number of ions per bunch. This success means that there is little scope for large gains in the future Pb-Pb luminosity. For this reason, collisions of other ion species are now being considered as part of the ALICE 3 detector upgrade proposal as a way to further increase nucleon-nucleon luminosity. 

In order to guide the design of the upgraded experimental detector and to choose the optimal ion species to collide, it is important to understand the potential bunch intensity and luminosity reach for a range of different ions. Apart from the short, low-intensity pilot run with Xe in 2017~\cite{schaumann18_ipac} and O and Ne in 2025~\cite{Bruce2021a}, there is no LHC experience with high intensity ions other than Pb. Other ions have, however, been accelerated in the CERN injector complex for fixed-target experiments~\cite{NA61_light_ions}. Although these beams were not optimised for LHC operation, they still provide important insights on beam production and acceleration. Some first estimates of the achievable luminosity for a range of different ions between O and Xe were presented in~\cite{YR_WG5_2018} based on a simple empirical power-law scaling of the bunch intensity $N_b$ from Pb beams to any ion species with a given mass number $A$ and atomic number $Z$:
\begin{equation}
\label{eq:ion_scaling_WG5}
N_b(Z, A) = N_b(82, 208) \Bigg( \frac{Z}{82}\Bigg)^{-p}.
\end{equation}
Here the fitting parameter $p$ is obtained from the limited experience with SPS fixed target beams, with $1.5 \geq p \geq 1.9$ used in~\cite{YR_WG5_2018} for luminosity projections of various ion species. This study indicates a potential high gain in integrated luminosity for the lightest ion species, in particular O. However, these studies did not account for realistic beam brightness limitations in the LHC injectors and were based on a simplified LHC luminosity model. Therefore, they now appear  too optimistic. 

In this article, we present predictions of the achievable LHC intensity and luminosity, taking limitations throughout the injector chain into account for a range of ion species, most of which the LHC has never accelerated. The detailed production scheme depends on electron cooling, bunch splitting, electron stripping and observed transmissions between machines, as well as other limitations of the injector chain. We account for space charge and electron cooling by scaling from experimentally achieved beam parameters. We start from a scheme similar to the present baseline Pb production scheme but also explore variations that could improve the intensity injected into the LHC. 

This paper is organized as follows: Sec.~\ref{sec:prodScheme} provides an overview of the production of ion beams in the CERN injector complex. Our development of the numerical \textbf{Injector Model} is presented in Sec.~\ref{sec:BaselineInt}, with the objective of realistically propagating bunch intensities through the injectors to LHC injection for any ion species. Relevant degrading effects such as space charge (SC), electron cooling, intra-beam scattering (IBS) and charge-changing beam-gas interactions are discussed in Secs.~\ref{sec:SC_limit}, \ref{sec:electron_cooling}, \ref{sec:IBS} and \ref{sec:beam-gas-interactions}. Bunch intensity predictions with the current \pb production (``baseline")  scheme are presented in Sec.~\ref{sec:baseline_predictions}. Section~\ref{sec:Improvements} identifies possible production scheme improvements and highlights estimated gains. These results are used in Sec.~\ref{sec:LHC} to estimate the achievable integrated luminosity in the LHC with detailed simulations, as done for \pb ~in~\cite{bruce21_EPJplus_ionLumi}. We outline three scenarios: \textbf{baseline}, assuming a beam production similar to Pb-Pb operation in Run~3 and Run~4, \textbf{optimistic}, incorporating further ideas whose operational feasibility remains to be studied, and \textbf{25~ns}, which assumes future hardware upgrades to achieve a shorter bunch spacing. For each scenario and ion type, we predict the integrated luminosity and compare with the present Pb-Pb scheme. Finally, Sec.~\ref{sec:conclusion} presents the conclusions on the achievable luminosity for various ion species, compares these to the previous studies in~\cite{YR_WG5_2018}, and suggests further studies to refine the predictions.

\section{The Injector Model}

Before constructing a realistic model of the ion injector chain, a detailed understanding of the current Pb beam production is imperative. Section~\ref{sec:prodScheme} introduces the Pb production steps, with Sec.~\ref{sec:BaselineInt} describing the baseline model algorithm and its ingredients.

\subsection{Present LHC lead ion beam production}
\label{sec:prodScheme}

\begin{figure}[!]
  \centering
  \includegraphics[width=\columnwidth]{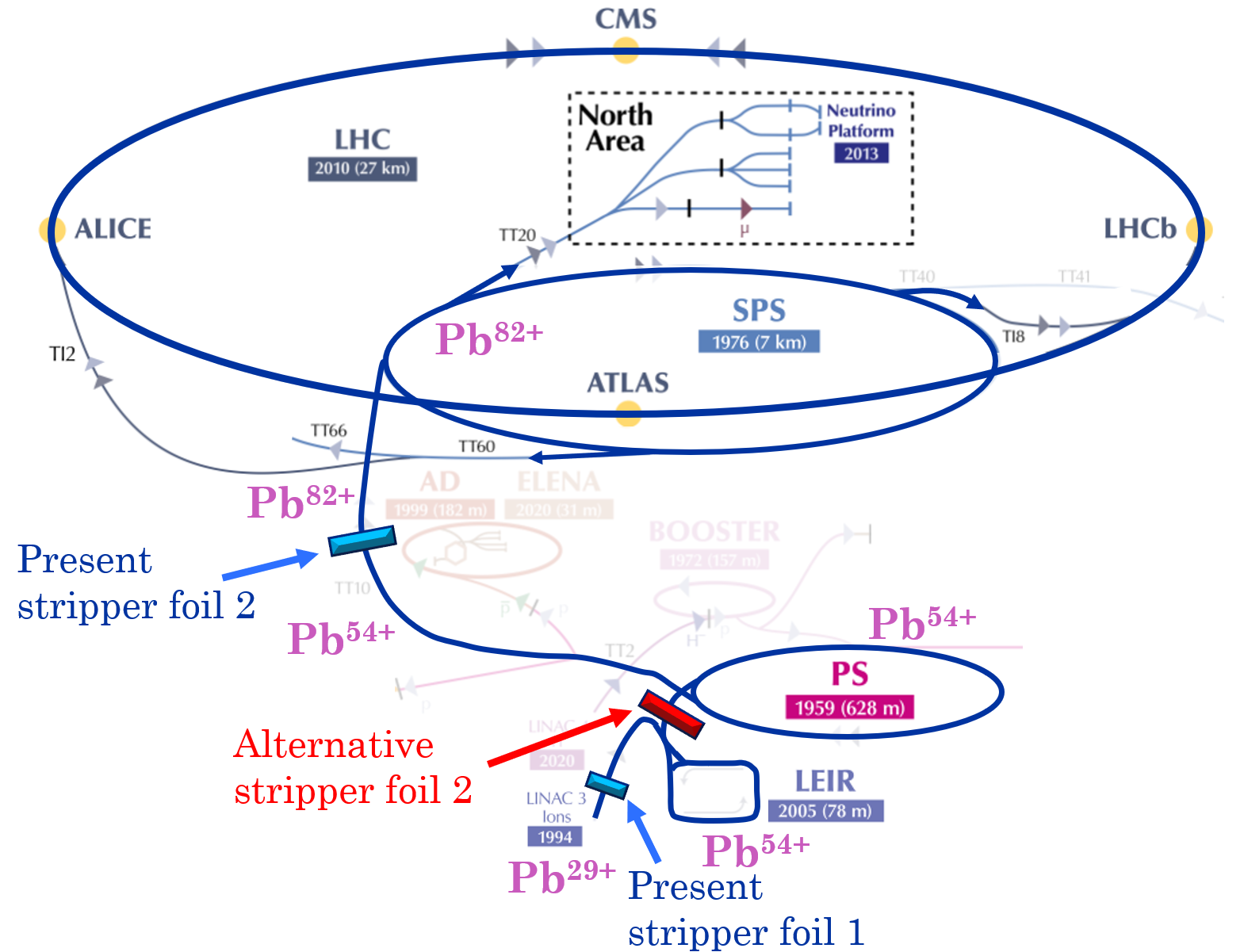}
  \caption{The Pb ion path, highlighted in dark blue, across the CERN injector chain and the LHC. Present and alternative stripper foil locations are shown, as well as the charge state evolution of Pb beams. Illustration projected on graphics from ~\cite{lopienska2022_cern_complex}.}
\label{fig:stripper_foils}
\end{figure}

\begin{figure}[t]
  \centering
  \includegraphics[width=.98\columnwidth]{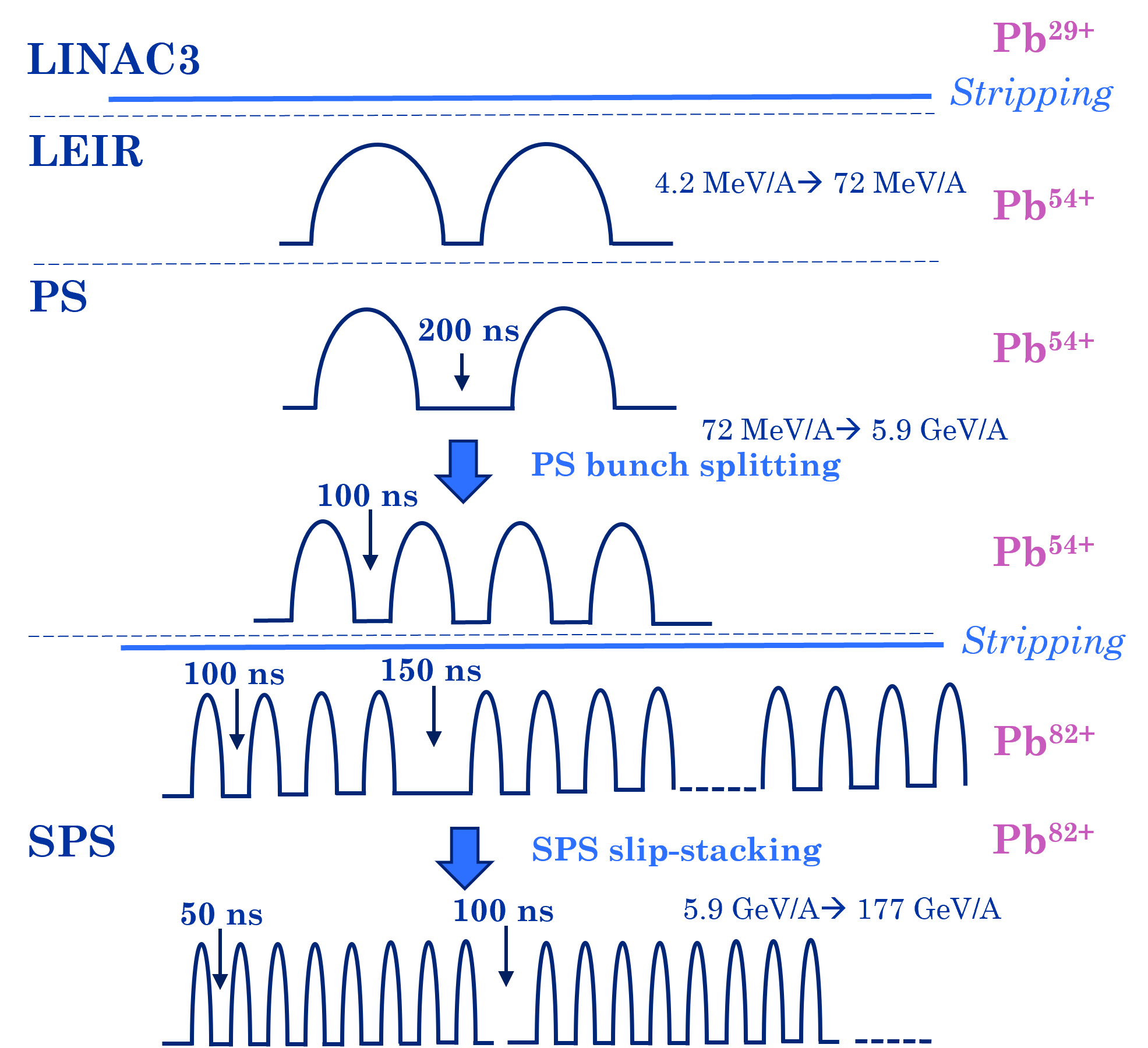}
  \caption{Present Pb ion beam production scheme and bunch train structure throughout the CERN ion injector chain.}
\label{fig:bunch_train_structure}
\end{figure}

The path of the Pb ions through the CERN accelerator complex is depicted in Fig.~\ref{fig:stripper_foils}, with the detailed production scheme for Pb beams described in~\cite{liu_TDR_ions}. Here we briefly review the aspects relevant to this article. The production of Pb ion beams, sketched in Fig.~\ref{fig:bunch_train_structure}, starts in the electron cyclotron resonance ion source (ECRIS), which provides partially stripped $^{208}\mathrm{Pb}^{29+}$ ions to a radio-frequency quadrupole and heavy-ion linac (Linac3) that accelerates them up to a kinetic energy of 4.2 MeV/nucleon. They are then stripped to $^{208}\mathrm{Pb}^{54+}$ and injected into the Low-Energy Ion Ring (LEIR). Several Linac3 pulses are accumulated in LEIR and transformed into two dense bunches, which are accelerated up to a kinetic energy of 72 MeV/nucleon. These two bunches are then transferred to the Proton Synchrotron (PS), where they are split into four bunches, accelerated to 5.9 GeV/nucleon and extracted with a bunch spacing of 100~ns~\cite{bartosik19_evian}. In the transfer line to the Super Proton Synchrotron (SPS), the remaining electrons are stripped off to achieve fully stripped \pb. Note that the splitting of the bunches in the PS also helps to reduce detrimental space charge effects in the SPS, as the density of particles per bunch is reduced.  

The final LHC train structure is defined in the SPS. For the baseline production scheme implemented in Run 3, pairs of four-bunch batches from the PS with 100~ns bunch spacing are interleaved through RF manipulations in the SPS to form eight-bunch batches with 50~ns bunch spacing, a process known as momentum slip-stacking~\cite{liu_TDR_ions}. The slip-stacking relied on an upgrade of the SPS RF system which was implemented during the second long LHC shutdown (2019-2022). Several pairs of four-bunch injections into the SPS can be accumulated to form longer trains of 8-bunch 50-ns batches with 100~ns spacing between the batches. The beam from the SPS is extracted at an energy of 177~GeV/nucleon (corresponding to $450\,Z$~GeV with fully stripped ions) and transferred to the LHC. Many such SPS trains made up of either 56 bunches and 40 bunches are accumulated in the LHC to form the full beam of up to 1240 bunches per LHC ring~\cite{bruce21_EPJplus_ionLumi}.

During Run~1 and Run~2, several LHC injector chain improvements increased the injected LHC bunch intensity over time~\cite{bartosik19_evian}. In 2018, an intensity of about $2.1\times10^8$~ions per bunch was reached~\cite{jowett19_evian}, which is three times higher than the original LHC design value of $7\times10^7$~ions per bunch. This was achieved with a different production scheme, with three bunches created in LEIR and no splitting in the PS, creating a train structure with 75~ns spacing in the LHC. During the 2023 LHC ion run, the 50~ns production scheme using SPS slip-stacking was commissioned and used for luminosity production for the first time. In 2024, optimizations along the injector complex allowed an intensity of $2.3\times10^8$~ions per bunch to be reached at the start of LHC collisions, surpassing the projected HL-LHC bunch intensity target of $1.8\times10^8$, at the price of slightly larger transverse emittance. This mode of operation is foreseen for the remainder of Run 3 and for Run~4. 

The principal remaining known intensity limitations for the Pb beams in the injector complex are due to SC and IBS. At high bunch intensities the SC tune shifts on the LEIR injection plateau can bring parts of the beam onto resonances, resulting in beam loss. A similar beam degradation from SC and IBS is observed on the SPS injection plateau. Such limitations in the PS have so far not caused any bottlenecks thanks to the fast energy ramp after injection, such that the SC tune shift decreases rapidly some tens of milliseconds after injection~\cite{waagaard_2023_HB}. 

\subsection{Injector model outline and parameters}
\label{sec:BaselineInt}

\begin{figure}[b]
  \centering
  \includegraphics[width=\columnwidth]{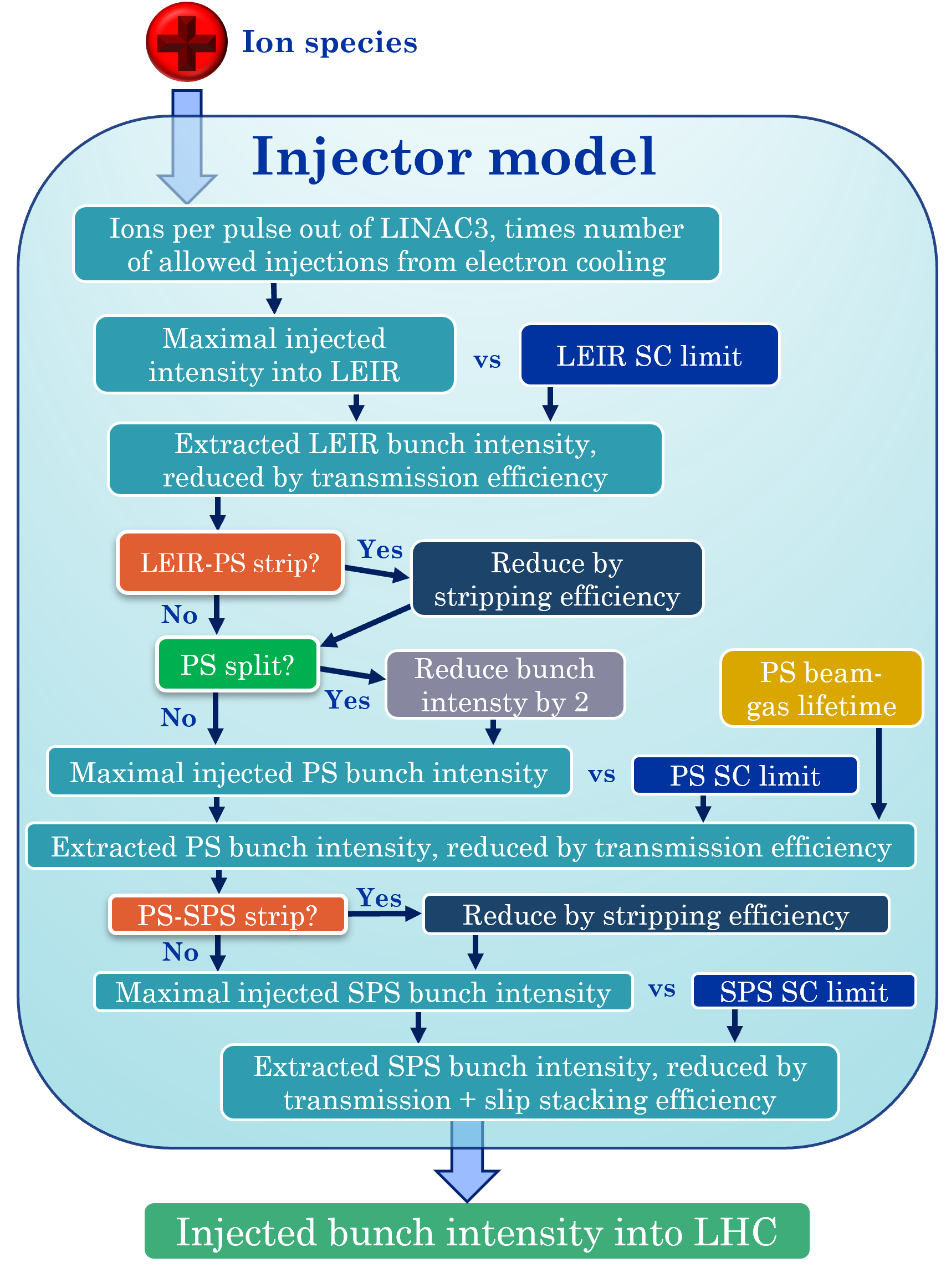}
  \caption{Flowchart of the  Injector Model.}
\label{fig:flowchart}
\end{figure}

In order to estimate the possible LHC bunch intensities for each ion species, we have implemented a numerical computer program, which we call the Injector Model, available at~\cite{injector_model_github_repo}. This propagates the beam parameters at injection and extraction in each machine across the injector chain: Linac3, LEIR, PS and SPS. The algorithmic flow is schematically illustrated in Fig.~\ref{fig:flowchart}. For a given ion species, this computes the maximum achievable injected intensity in each machine, taking transmission and injection inefficiencies into account based on experimental Pb values. The Injector Model also includes a ``space charge limit" in each machine, discussed in detail in Sec.~\ref{sec:SC_limit}, which is the highest intensity that can be propagated to the next machine. Beam lifetime reductions from interactions with rest gas in the PS are also considered. Further intensity reductions come from IBS and from the stripping stages between Linac3 and LEIR, as well as between the PS and the SPS, where all remaining electrons are removed. 

As a starting point, we assume a similar production scheme in the injectors as for present Pb operation and discuss possible improvements in Sec.~\ref{sec:Improvements}. 
\begin{table}[t]
\centering
\caption{Properties of considered ion species, their assumed charge state and Linac3 current. Values from~\cite{Scrivens_gamma_factory_source_status} are marked by $\dag$, from \cite{detlef_Kr_source} by $\ddag$, from \cite{detlef_Mg_source} by $\star$, with those marked $\diamond$ achieved in 2023. Untested currents are indicated by $\odot$, for which we assume 25 $\mu$A, the lowest current experimentally achieved among the heavier ions.}
\begin{adjustbox}{width=.85\columnwidth}
\begin{tabular}{rrrrrr} \hline
Ion & $Z$ & $A$ & charge in         & mass & Linac3  \\ 
    &     &     & LEIR-PS           &    [GeV/$c^2$]        & current [$\mu$A] \\   \hline 
He  & 2   & 4   & 1                 & 3.73      & 160$^{\dag}$   \\  
O   & 8   & 16  & 4                 & 14.86     & 90$^{\diamond}$      \\ 
Mg  & 12  &  24 & 7                 & 22.36     & 20$^{\star}$     \\  
Ar  & 18  & 40  & 11                & 37.22     & 60$^{\dag}$     \\  
Ca  & 20  & 40  & 17                & 37.22     & 25$^{\odot}$   \\  
Kr  & 36  & 86  & 22 (29)                & 80.03     & 40$^{\ddag}$ (25$^{\odot}$)     \\  
In  & 49  & 115 & 37                & 107.01    & 25$^{\dag}$    \\  
Xe  & 54  & 129 & 39                & 120.05    & 30$^{\dag}$      \\  
Pb  & 82  & 208 & 54                & 193.69    & 30$^{\dag}$    \\ \hline
\end{tabular}
\end{adjustbox}
\label{tab:species_general}
\end{table}
The ion species of interest are presented in Table~\ref{tab:species_general}.  The estimated ion currents from Linac3 are based upon previously achieved values during operational ion production tests~\cite{Scrivens_gamma_factory_source_status}. Other charge states than default values in Table~\ref{tab:species_general} may be possible from a production point of view in the source and Linac3, but may yield lower intensities in LEIR because of space charge, for example O$^{5+}$. Charge state optimisation will be discussed in detail in Section~\ref{sec:charge_state_and_isotope_scan}. Most of these ions were previously studied in Ref.~\cite{YR_WG5_2018}. This study considers ions of interest for LHC experiments up to lead. Heavier elements are excluded due to radiological, chemical, and operational hazards and incompatibility with the current ion source. 


Some ions have been used previously for SPS fixed target experiments (Ar in 2015, and Xe in 2017), while In was used back in 2003. Xe was also collided in the LHC in 2017~\cite{schaumann18_ipac, fuster18_ipac_Xe}. Oxygen beams were used to set up LEIR for the first time in 2005. Calcium has been proposed for a Gamma-Factory design~\cite{krasny2015gamma,arek20_PSI_collimation}, which relies on partially-stripped heavy-ion beams in the LHC as the basis for a novel type of light source. For Ca, which has never been used in the CERN complex, we conservatively assume the lowest current among the heavier operationally tested species. 

The injector parameters used are shown in Table~\ref{tab:injector_parameters}. To fill LEIR, we consider up to 8~possible injections from Linac3 with a 300~$\mu$s pulse length, as used for the Pb beam production in 2024~\cite{slupecki_iefc_2024_leir_performance}. The number of ions before capture in LEIR quoted in this table considers a constant injection efficiency of 50\% and no losses in between injections. In reality, further limitations from decreasing injection efficiencies along the injection plateau as well as losses during the accumulation process may further limit the total accumulated intensity in LEIR. For example, the effective transmission (including varying injection efficiencies and losses on the injection plateau) for Pb ions is about 70\%. For O ions, an effective transmission on the injection plateau of more than 90\% was measured in the pilot run in 2025. These effects need to be addressed in more detail in future studies. With the presently available knowledge, these fluctuations combined with transmission uncertainties across the injectors lead us to estimate uncertainties in injected LHC intensity of around 20-30\%.  

In the baseline scenario, we assume that all the presently observed losses in the PS-SPS transfer line are due to the stripper foil. We also assume similar PS-SPS stripping efficiencies for all ion species and conservatively use the value listed in Table~\ref{tab:injector_parameters}  unless stated otherwise. Furthermore, we account for the possible splitting of bunches in LEIR and in the PS, where each splitting halves the bunch intensity. In the PS, no obvious intensity limitations have been encountered for the present Pb beams~\cite{bartosik19_evian}.

\begin{centering}
\begin{table}[t]
  \caption{Assumed injector parameters, based on operational experience with Pb~\cite{2019_sps_recent_Pb_performance} and LIU ion beam parameters~\cite{LIU_ion_targets}.}
 \begin{adjustbox}{width=0.85\columnwidth}
\begin{tabular}{lr} \hline
                             & Standard scheme      \\ \hline
Max n.o.~Linac3 pulses       & 8                    \\
LEIR injection efficiency    & 50\%                  \\
N.o. bunches in LEIR            & 2                     \\
LEIR transmission efficiency & 76\%                   \\
LEIR-PS transmission efficiency & 93\%                   \\
PS transmission efficiency & 92\%                   \\
N.o. extracted bunches from PS              & 4                  \\
PS $B_{\textrm{min}}$ & 0.0383 T \\
PS $B_{\textrm{max}}$ & 1.26 T \\
Stripping efficiency PS-SPS  & 90\% \\
SPS transmission efficiency  & 72\% \\
\hline
\end{tabular}
 \end{adjustbox}
 \label{tab:injector_parameters}
\end{table}
\end{centering}

For each ion species, the Injector Model calculates the momentum $p$, kinetic energy per nucleon and the relativistic gamma factor $\gamma$ at injection and extraction from each machine, considering the magnetic rigidity 
\begin{equation}
B \rho = \frac{p}{q}
\label{eq:Brho}
\end{equation}
of each accelerator, with $q$ being the charge. If the charge state $q$ is unchanged, $B \rho$ at extraction of one accelerator is identical to $B \rho$ at injection for the next accelerator. If stripping occurs, leading to a change in charge state, a new $B \rho$ is calculated from Eq.~\eqref{eq:Brho} but the relativistic $\gamma$ remains unchanged. At extraction of Linac3, a fixed $\gamma$ is assumed, given by the Linac3 accelerating structures. The proton-equivalent momentum $p/q$ is also calculated to see if we inject at too low a momentum for the accelerator's dipole magnets to handle, in particular for the PS, where the dipole magnetic field ramps from a minimum of $B_{\textrm{min}} = 0.0383$ T to a maximum field of $B_{\textrm{max}} = 1.26$ T. 

It is worth noting that the Injector Model can also be used to estimate the intensity available for fixed target experiments in the PS or SPS Experimental Areas, such as for for the lighter ions requested by NA61/SHINE~\cite{north_area_requests}. 

\subsection{Space charge limit}
\label{sec:SC_limit}

We define the space-charge limit (SCL) for a given ion species as the bunch intensity at which space-charge forces become equivalent to those of Pb beams, which currently limits the ion operational intensity, primarily in LEIR and SPS. Exhaustive optimisation work has already been accomplished throughout the complex in the present configuration~\cite{bartosik19_evian}, although we do not exclude future incremental improvements for Pb. We start from the expression for the SC tune shift $\Delta Q_u$, calculated as an integral over the lattice elements around the circumference of the ring~\cite{schindl_space_1999}:
\begin{equation}
 \Delta Q_u = - \frac{r_0 \hat{\lambda}}{2\pi 
 \gamma (\gamma^2-1)
 }
 \oint \frac{\beta_u(s)}{\sigma_u(s)\left(\sigma_x(s)+\sigma_y(s)\right)}\mathrm{d}s,
 \label{eq:SCfull}
\end{equation}
where $u$ denotes either transverse plane, $x$ or $y$, and $\beta_u$ and $\sigma_u$ are the betatron function and the beam size in plane $u$ respectively. Furthermore, $r_0=(q e)^2/(4\pi \epsilon_0 m c^2)$ is the classical particle radius, with $q$ and $m$ being the charge state and mass of the ion, $e$ the elementary charge and $\epsilon_0$ the vacuum permittivity. Finally, $\hat{\lambda}=N_b/(\sqrt{2\pi}\sigma_z)$ is the peak ion line density, $N_b$ the bunch population and $\sigma_z$ the rms bunch length. 

The limiting space-charge tune-shift  $\Delta Q_{u, \mathrm{Pb}}$  in each injector can be calculated from Eq.~\eqref{eq:SCfull} and the operationally achieved intensity $N_{b, \mathrm{Pb}}$ with Pb bunches.  
To calculate the maximum achievable $N_b$ for another ion, we solve Eq.~\eqref{eq:SCfull} for $N_b$, while assuming that the SC tune shift cannot be larger than the $\Delta Q_{u, \mathrm{Pb}}$ that was achieved with Pb:  
\begin{equation}
 N_{b} = \frac{-2 \pi \sqrt{2 \pi} 
 \gamma (\gamma^2-1)
 \sigma_z 
 \Delta Q_{u, \mathrm{Pb}}}{r_{0} e \oint \frac{\beta_u(s)}{\sigma_u(s)\left(\sigma_x(s)+\sigma_y(s)\right)}\mathrm{d}s}.
 \label{eq:SC_limit_full}
\end{equation}

In a second step, we assume that the dispersive contribution to $\sigma_u$ is small compared to the betatronic one, such that $\sigma_u\approx\sqrt{\epsilon_{u}\beta_u(s)}$, with $\varepsilon_{u}$ being the geometric transverse emittance. We additionally assume that the beams are round ($\varepsilon_{x}=\varepsilon_{y}$), that the optics configurations do not change across the ion species and that all ion species have similar geometric emittances. Eq.~\eqref{eq:SC_limit_full} then simplifies into a new expression for the SC-limited bunch intensity of the new ion species under consideration as compared to Pb

\begin{equation}
 N_{b}  \approx N_{b,\mathrm{Pb}}\frac{q_\mathrm{Pb}^2 m\gamma(\gamma^2-1)}{q^2 m_\mathrm{Pb} \gamma_\mathrm{Pb}(\gamma_\mathrm{Pb}^2-1) }.
 \label{eq:SC_limit}
\end{equation}

Assuming identical $\varepsilon_{x,y}$ and $\sigma_z$ (from Pb operation) for all ion types, Eq.~\eqref{eq:SC_limit} rather than Eq.~\eqref{eq:SC_limit_full} can be used in the calculations.
We further assume that beam losses due to space charge depend only on the machine properties, for instance non-linearities and beam behaviour at different high-order resonances, and not on the particle type.

\subsection{Electron cooling}
\label{sec:electron_cooling}

Electron cooling is a technique to decrease the emittance of a given beam, widely implemented in accelerators worldwide, in particular at lower beam energies. A detailed theoretical overview of electron cooling is given in Ref.~\cite{parkhomchuk_overview}. One of the main figures of merit is the cooling time $\tau_{\mathrm{cool}}$, usually defined as the time to decrease the beam emittance by a factor $1/e$, i.e., $\varepsilon(\tau) = \varepsilon(0) /e$ (with the Euler number $e = 2.71828$). The cooling time increases strongly with beam energy and decreases with higher charge states. The LEIR electron cooler was designed for stacking Pb$^{54+}$ ions injected from Linac3. The number of injections from Linac3 into LEIR is limited by how quickly each pulse can be cooled before the next injection. Knowledge of electron cooling times $\tau_{\mathrm{cool}}$ for other ion species is therefore crucial in estimating the appropriate number of injections into LEIR.

The LEIR electron cooler performance with future ion species has been studied with a numerical model based on the Parkhomchuk formalism~\cite{parkhomchuk2000_model}, implemented in Xsuite~\cite{xsuite} and recently benchmarked against BETACOOL~\cite{kruyt_xsuite_vs_betacool}. 
The main findings are summarised in Table~\ref{tab:cooling_time_ion_future_species}. 
For each of the considered ion species and their assumed LEIR charge state, it gives the computed cooling time $\tau_{\mathrm{cool}}$. Assuming that the LEIR electron cooler is presently operated at its maximum performance with Pb$^{54+}$ beams, the relative cooling time compared to Pb indicates the number of Linac3 injections possible for other ion species. The rightmost column of Table~\ref{tab:cooling_time_ion_future_species} shows the estimated number of allowed Linac3 injections considering cooling performance and available cooling time on the LEIR injection plateau. He$^{2+}$, O$^{4+}$, Mg$^{7+}$ and Ar$^{11+}$ will be limited to a single Linac3 injection, while Ca$^{17+}$ can profit from up to four injections, Kr$^{22+}$ up to three, In$^{37+}$ up to seven and Xe$^{39+}$ up to six. The number of Linac3 injections considered in the Injector Model for each ion species is the maximum allowed. Note that even a single injection may be difficult to cool for He and O ions, which could result in losses and thus limit the achievable intensity. Single-bunch O$^{4+}$ pilot beam commissioning in LEIR in 2025 did indeed reveal difficulties in properly cooling more than one Linac3 injection for such oxygen beams.
\begin{table}[t]
\centering
\caption{Estimated cooling performance in LEIR for the ions studied in this paper. The cooling times are given as both absolute values and relative to the case of Pb. The last column indicates the possible number of Linac3 injections into LEIR, considering the cooling performance and available cooling time.}
\begin{adjustbox}{width=.98\columnwidth}
\begin{tabular}{rrrrrrr} \hline
Ion & $Z$ & $A$ & charge in        & Cooling & Relative & N.o. LEIR \\ 
    &     &     & LEIR-PS           & time $\tau_{\mathrm{cool}}$ [s] & cooling time & injections \\   \hline 
He  & 2   & 4   & 1                    & 8.69  & 48.8 & 1 \\  
O   & 8   & 16  & 4                    & 2.01  & 11.3 & 1  \\ 
Mg  & 12  &  24 & 7                    & 1.05  & 5.9  & 1  \\  
Ar  & 18  & 40  & 11                   & 0.73  & 4.1  & 1 \\  
Ca  & 20  & 40  & 17                   & 0.33  & 1.9  & 4\\  
Kr  & 36  & 86  & 22                   & 0.42  & 2.4  & 3\\  
In  & 49  & 115 & 37                   & 0.20  & 1.1  & 7 \\  
Xe  & 54  & 129 & 39                   & 0.21  & 1.2  & 6 \\  
Pb  & 82  & 208 & 54                   & 0.18  & 1.0  & 8 \\ \hline
\end{tabular}
\end{adjustbox}
\label{tab:cooling_time_ion_future_species}
\end{table}

The scaling law $\tau_{\mathrm{cool}} \propto A/q^2$ can be used to estimate the cooling time of other charge states for a given ion species. In general, 
the cooling time $\tau_{\mathrm{cool}}$ of a charge state $q$ and mass number $A$ is scaled from Pb as
\begin{equation}
\label{eq:ecooling_scaling}
\tau_{\mathrm{cool}} =
\tau_{\mathrm{cool \text{, Pb}}} \times \bigg( \frac{q_{\text{Pb}}}{q} \bigg)^2 \frac{A}{A_{\text{Pb}}}.
\end{equation}

Experience with lighter ions in the LEIR electron cooler is scarce. Aside from a brief O$^{4+}$ beam test during the 2005 commissioning~\cite{leir_ecooler}, a test of Mg$^{7+}$ in 2024 demanded considerable effort to configure the electron cooler, despite being only a single-injection, low-intensity beam. 


\subsection{Intra-beam scattering}
\label{sec:IBS}

IBS is a stochastic process of small-angle multiple Coulomb scattering of charged particles in a beam, leading to momentum exchange between the planes and emittance growth. This beam-degrading effect is particularly important for dense, low-energy ion beams with high charge states, but is also visible in proton beams stored for many hours in high-energy storage rings such as the LHC~\cite{papadopoulou_bunch_2020}. Many IBS formalisms exist, including Piwinski~\cite{piwinski_intra-beam-scattering_1974}, Bjorken-Mtingwa (BM) for strong-focusing lattices~\cite{bjorken_intrabeam_1983}, Nagaitsev~\cite{nagaitsev_intrabeam_2005} for faster numerical BM elliptic integral evaluation, and the kinetic formalism by Zenkevich~\cite{zenkevich_new_2006}. A way to quantify the IBS strength is through emittance growth rates $1/\tau_{u}$, with $u = x, y, z$. This parameter indicates the IBS strength in the respective planes for a given set of beam parameters. The IBS growth rates depend on the machine optics, the charge state, the bunch intensity $N_b$, the geometric emittance $\varepsilon_u$ and the beam energy.

The Nagaitsev formalism model implementation in Xsuite~\cite{xsuite, felix_ibs_study_2024}  was used to numerically compute the IBS growth rates for all considered ion species in LEIR, PS and SPS, assuming ion beam parameters values of $\varepsilon_{x,y}^n$, $\delta$ and $\sigma$ from~\cite{John2021}. The bunch intensity $N_b$ was chosen for each ion species to be at its SC limit from Eq.~\eqref{eq:SC_limit}. Although not all ions reach the SC limit in each machine, this intensity value provides a fair comparison between species, when SC effects start coming into play and intensity limits from collective effects arise. The computed relative IBS growth rates, the absolute growth rates divided by the Pb growth rates, at injection energy are shown in Table~\ref{table:IBS_rel_growth_rates}. 

\begin{table}[t]
\caption{Relative growth rates $T_\textrm{rel} = (1/\tau_u)/(1/\tau_{u, \mathrm{ Pb}})$: ratio of analytical Nagaitsev IBS growth rates and the corresponding Pb values at injection energy.}
\label{table:IBS_rel_growth_rates}
\begin{adjustbox}{width=.98\columnwidth}
\begin{tabular}{llllllllll}
$T_\textrm{rel}$ & He   & O    & Mg   & Ar   & Ca    & Kr   & In   & Xe   & Pb \\ \hline \hline
$x$, LEIR  & 0.02 & 0.08 & 0.17 & 0.24 & 0.48 & 0.43 & 0.86 & 0.85 & 1  \\
$y$, LEIR & 0.02 & 0.08 & 0.17 & 0.24 & 0.55 & 0.42 & 0.86 & 0.85 & 1  \\
$z$, LEIR & 0.02 & 0.08 & 0.17 & 0.24 & 0.39 & 0.43 & 0.86 & 0.85 & 1  \\ \hline
$x$, PS   & 0.02 & 0.08 & 0.19 & 0.26 & 0.53 & 0.42 & 1.09 & 1.02 & 1  \\
$y$, PS   & 0.02 & 0.08 & 0.17 & 0.24 & 0.62 & 0.42 & 0.95 & 0.92 & 1  \\
$z$, PS   & 0.02 & 0.08 & 0.16 & 0.24 & 0.64 & 0.42 & 0.87 & 0.86 & 1  \\ \hline
$x$, SPS  & 0.04 & 0.15 & 0.17 & 0.25 & 0.34 & 0.5  & 0.47 & 0.56 & 1  \\
$y$, SPS  & 0.04 & 0.14 & 0.18 & 0.26 & 0.26 & 0.49 & 0.54 & 0.62 & 1  \\
$z$, SPS & 0.04 & 0.16 & 0.11 & 0.21 & 0.22 & 0.51 & 0.03 & 0.26 & 1 \\ \hline
\end{tabular}
\end{adjustbox}
\end{table}

In the present Pb production scheme, IBS is assumed to have the largest impact on the SPS injection plateau, which lasts for about 45 s with 14 injections. In the transverse planes, Kr, In and Xe have growth rates of around half that of Pb, while for the lighter ions the growth rates are  much smaller. Hence, we conclude that IBS will most likely be a second-order effect in terms of losses for most light ions, and it is therefore not directly included in the Injector model at this stage. Future studies will address more detailed emittance evolution models including IBS through the ion injector chain, which may affect beam quality differently for different ion species.


\subsection{Beam interactions with rest gas}
\label{sec:beam-gas-interactions}

Charge-exchanging beam-gas interactions can occur when particles lose or capture electrons from residual gas molecules in the vacuum pipe. These processes can cause beam losses when the particles that now have a different charge state fall outside the machine acceptance. The two dominant atomic interaction effects between the beam and the residual gas in the CERN ion injector chain are electron capture (EC) and electron loss (EL), which can involve multiple capture and loss processes with many electrons~\cite{basic_atomic_interactions_book_tolstikhina2018}. EC cross sections dominate at lower energies, but decrease rapidly at higher energies, where EL is the main charge-state-changing process. Other beam-gas interaction processes, such as inelastic nuclear collisions and elastic nuclear Coulomb scattering exist, but typically have much smaller cross sections than EC and EL in this context. The lifetime $\tau_i$ due to these atomic charge-changing interactions with a residual gas component $i$ is given by
\begin{equation}
\tau_i = \frac{1}{\sigma\,n_i\,\beta \, c },
\label{eq:beam_lifetime}
\end{equation}
where $\sigma = \sigma_{\textrm{EC}} + \sigma_{\textrm{EL}}$ is the total charge-changing cross section, $\beta$ is the projectile relativistic beta factor, $c$ is the speed of light and $n$ is the molecular number density in the beam pipe. The total lifetime $\tau$ is calculated by inversely adding $\tau_i$ for all known residual gas components and partial pressures. If vacuum conditions, projectile energies and cross sections are known, Eq.~\eqref{eq:beam_lifetime} can be used to predict ion beam lifetimes in storage rings due to charge-changing electron loss and capture mechanisms. Application of such predictions include calculation of heavy-ion beam lifetime estimates for FAIR~\cite{shevelko2018lifetimes_ricode_m} and estimating the dynamic vacuum requirements for LEIR~\cite{mahner2007_LEIR_vacuum_system}. In the context of the CERN ion injector chain, the few systematic studies on this topic include the 2017 SPS pilot study of $^{129}$Xe$^{39+}$ ions, which concluded that beam losses were driven predominantly by beam-gas interactions~\cite{hirlaender_xe_lifetime_2018}. 


For each ion species in each machine, the cross sections are calculated with the open-access Python repository \verb|beam_gas_collisions|~\cite{beam_gas_collisions_github_repo}, which uses the established Schlachter formula~\cite{schlachter_electron_1983} for $\sigma_{\textrm{EC}}$ and the Weber semi-empirical formula for total $\sigma_{\textrm{EL}}$~\cite{weber_2016_semi_empirical_formula}, adjusted to more than 100 experimental data points of various projectiles at different energies with different target atomic numbers. The estimated static vacuum conditions with average pressure values $\bar{P}$ and rest gas composition in LEIR, PS and SPS are shown in Table~\ref{table:gas_fractions}. These assumed static pressure values, in particular the residual gas composition, are \textit{highly uncertain} and may be influenced by processes such as dynamic material desorption from lost ion particles. Thus the beam lifetimes computed with Eq.~\eqref{eq:beam_lifetime} due to EC and EL should be considered \textit{approximate}. 

\begin{table}[t]
\centering
\caption{Assumed average pressure values $\bar{P}$ and rest gas composition (fraction of total particle density). The LEIR vacuum system is discussed in Ref.~\cite{mahner2007_LEIR_vacuum_system}, and the SPS vacuum conditions are obtained from~\cite{hirlaender_xe_lifetime_2018}. The LEIR and PS base pressures represent typical gauge measurements from Nov. 2024. The rest gas composition is assumed to be dominated by H$_2$, with other gas fractions very approximate.}
\label{table:gas_fractions}
\begin{adjustbox}{width=.7\columnwidth}
\begin{tabular}{c|ccc} 
          & LEIR       & PS                   & SPS       \\ \hline
          & & & \\[-2.2ex]
$\bar{P}$ {[}mbar{]} & $10^{-11}$ & $1.2 \times 10^{-9}$ & $10^{-8}$ \\ \hline 
H$_2$ fraction           & 0.83       & 0.9                  & 0.905     \\
H$_2$O fraction          & 0.02       & 0.1                  & 0.035     \\
CO fraction          & 0.04       & 0                    & 0.025     \\
CH$_4$ fraction          & 0.05       & 0                    & 0.025     \\
CO$_2$ fraction          & 0.06       & 0                    & 0.01  \\ \hline 
\end{tabular}
\end{adjustbox}
\end{table}

\begin{table}[t]
\centering
\caption{Calculated beam lifetime $\tau$ (in seconds) due to EL and EC in LEIR (injection) and the PS (injection and extraction) for partially stripped ions, used to compute the PS beam transmission $\eta_{\text{PS}}$.}
\label{table:full_EC_EL_lifetime_predictions}
\begin{adjustbox}{width=.85\columnwidth}
\begin{tabular}{l|llll}
$q_{\mathrm{LEIR,PS}}$     & $\tau_{\text{LEIR, inj}}$ [s] & $\tau_{\text{PS, inj}}$ [s]   &  $\tau_{\text{PS, extr}}$ [s] & $\eta_{\text{PS}}$\\ \hline \hline
He$^{1+}$  & 160 & 4.0                 & 5.4 & 76\% \\
O$^{4+}$   & 130 & 3.3                 & 4.6 & 73\% \\
O$^{5+}$   & 120 & 2.8                 & 3.8 & 69\% \\
Mg$^{7+}$  & 230 & 5.5                 & 7.7 & 82\% \\
Ar$^{11+}$ &  370 & 9.1                 & 11.9 & 88\% \\
Ca$^{17+}$ & 610 & 25  & 26 & 92\% \\
Kr$^{22+}$ & 280 & 12  & 17 & 91\% \\
In$^{37+}$ & 150 & 43 & 54 & 92\% \\
Xe$^{39+}$ & 140 & 36  & 48 & 92\% \\
Pb$^{54+}$ & 69 & 39 & 47 & 92\% \\ \hline        
\end{tabular}
\end{adjustbox}
\end{table}

Estimated lifetimes due to electron loss and electron capture in LEIR and the PS are shown in Table~\ref{table:full_EC_EL_lifetime_predictions}. These effects are negligible in the SPS, where ions are fully stripped, energies are high and only electron capture is possible. Fully stripped ions such as He$^{2+}$ and O$^{8+}$ also have long lifetimes in LEIR and the PS as they are 
are only marginally effected by electron capture and electron loss is not possible. In LEIR, which has the lowest base pressure of all machines, all computed lifetimes exceed the typical operational single injection cycle length (around 3 s) with a large margin. Beam-gas interactions in the PS are a potential bottleneck for some partially stripped lighter ion species such as He$^{1+}$, O$^{4+}$, O$^{5+}$ and Mg$^{7+}$, where beam transmission can be significantly worsened due to these effects. For O and He, this problem can be avoided by injecting fully stripped ions, but their higher kinetic projectile energy (compared to partially stripped ions) may cause radiation protection issues in LEIR. 

Due to the long beam lifetimes in LEIR, as seen in Table~\ref{table:full_EC_EL_lifetime_predictions}, we only consider transmission reductions from beam-gas interactions in the PS at this stage. To account for this reduced PS transmission due to beam-gas interactions in the Injector Model, the average beam lifetime over the cycle $\tau_{\text{PS, EC+EL}} = (\tau_{\text{PS, inj}} + \tau_{\text{PS, extr}})/2$ from electron capture and electron loss is calculated for the given charge state using \verb|beam_gas_collisions|, with Table~\ref{table:full_EC_EL_lifetime_predictions} showing the lifetimes for the most relevant ion charge states. The total PS lifetime $\tau_{\text{PS, total}}$ combines these lifetimes inversely: $1/\tau_{\text{PS, total}} = 1/\tau_{\text{PS, EC+EL}} + 1/\tau_{\text{PS, other}}$, where $\tau_{\text{PS, other}}$ corresponds to around 3\% losses from unknown sources not explained by beam-gas interactions, as observed for Pb$^{54+}$. The PS transmission is expressed as $\eta_{\text{PS}} = \exp(-t_{\text{PS, cycle}}/\tau_{\text{PS, total}})$, where $t_{\text{PS, cycle}} = 1.2$ s represents the typical length the beam spends in the accelerator. An upper limit of 92\% for PS transmission is still assumed, as in Table~\ref{tab:injector_parameters}. These beam-gas loss factors are more critical for lighter, partially stripped ion species such as O$^{4+}$ and Mg$^{7+}$. No major beam transmission issues from electron loss or electron capture are expected for the heavier proposed future ion species. 

\begin{figure}[t]
  \centering
  \includegraphics[width=.96\columnwidth]{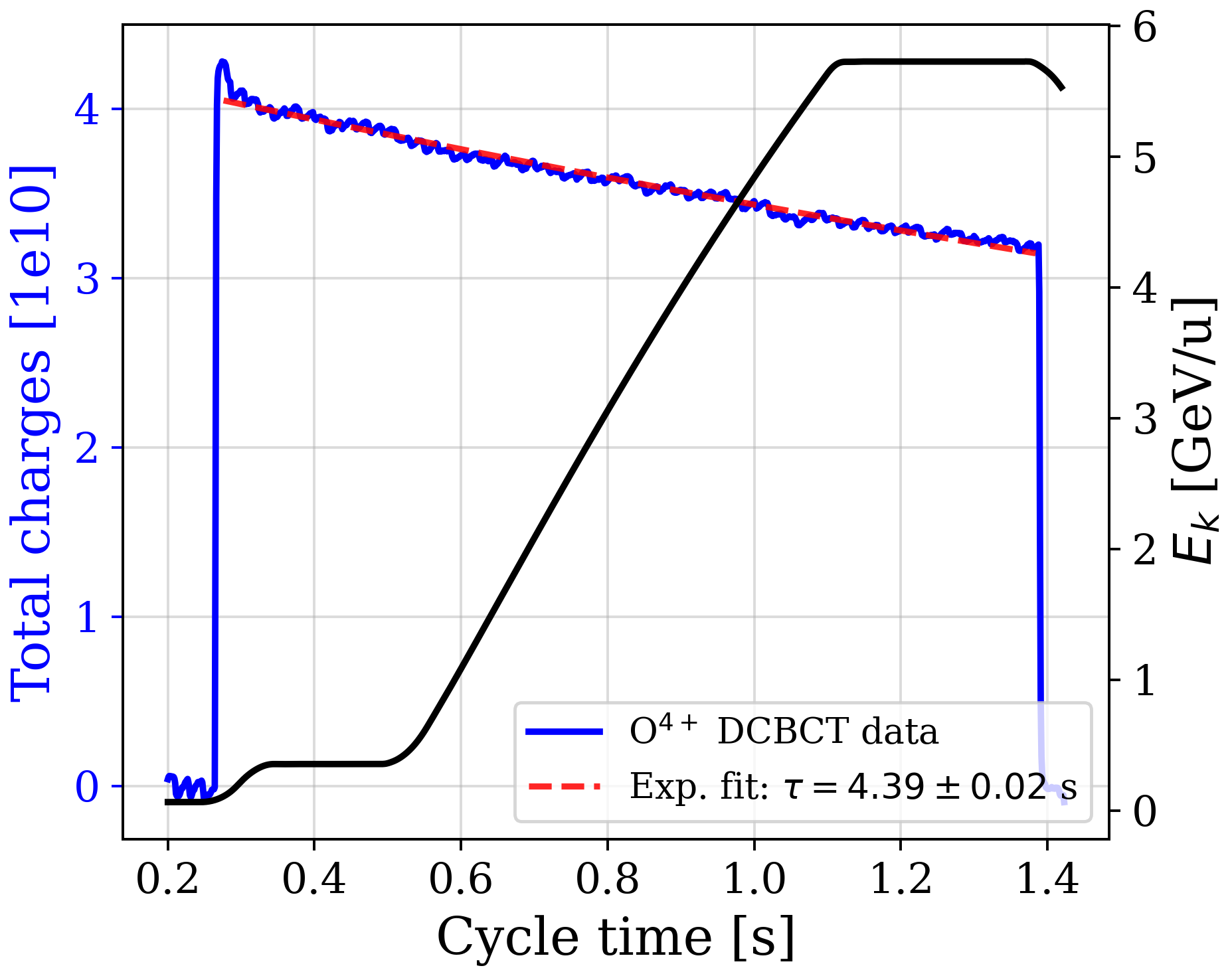}
  \caption{Example of measured intensity for single-bunch O$^{4+}$ pilot beams in the PS, from June 10$^{\text{th}}$ 2025.}
\label{fig:PS_O4_BCT_and_fit_2025_06_paper}
\end{figure}

During the June and July 2025 single-bunch O$^{4+}$ pilot beam commissioning, beam lifetimes of 4.4 seconds were measured in the PS, as shown in Fig.~\ref{fig:PS_O4_BCT_and_fit_2025_06_paper}. This value falls within the calculated lifetime at injection and extraction in Table~\ref{table:full_EC_EL_lifetime_predictions}. However, it should be noted that the measured O$^{4+}$ transmission varied strongly according to the PS average pressure, which in turn depends on regular sublimation pumping of the vacuum. Without sublimation pumping, beam transmission was only around 50\%. This improved to above 70\% within the first three hours after sublimation, similar to the calculated O$^{4+}$ transmission in Table~\ref{table:full_EC_EL_lifetime_predictions}. Operation with such lighter ion species will therefore require more detailed PS vacuum studies and improvements. 

\section{Injector model results}

\subsection{Baseline scenario predictions}
\label{sec:baseline_predictions}

\begin{table*}[ht!]
\caption{Baseline ion production scenario: predicted theoretical maximum bunch intensities $N_b$ (in number of ions) and space charge limit (SCL) at the injection energies along the injector chain. In LEIR, $N_b$ represents the total intensity at injection before capture. Two values are shown for $N_b$: one for \textcolor{blue}{8 injections} in LEIR (achieved for Pb) and one assuming the Electron-Cooling Limit (\textcolor{BrickRed}{ECL}) from Table~\ref{tab:cooling_time_ion_future_species}. For each ion species, the \underline{\textbf{global bottleneck}} in LEIR, PS or SPS is marked in bold and underlined.} 
\label{tab:results_baseline}
\begin{adjustbox}{width=\textwidth}
\centering
\begin{tabular}{cccc|cc|cc|cc} \hline
   & \multicolumn{3}{c}{LEIR injection (accumulated intensity before capture)}                                                                                                              & \multicolumn{2}{c}{PS injection}                                                                                                                            & \multicolumn{2}{c}{SPS injection}                                                                                                                           & \multicolumn{2}{c}{LHC injection}                                                                                                                \\ \hline
    & \begin{tabular}[c]{@{}c@{}}N.o. LEIR\\ injections\end{tabular}
   & \begin{tabular}[c]{@{}c@{}}Max $N_b$ \\ \textcolor{blue}{8 inj}, \textcolor{BrickRed}{ECL}\end{tabular} & \begin{tabular}[c]{@{}c@{}} SCL \\\end{tabular} & \begin{tabular}[c]{@{}c@{}}Max $N_b$, \\ with bunch splitting \end{tabular} & \begin{tabular}[c]{@{}c@{}}SCL \\ (at least) \end{tabular} & \begin{tabular}[c]{@{}c@{}}Max $N_b$ \\ \end{tabular} & \begin{tabular}[c]{@{}c@{}}SCL\\ \end{tabular} & \begin{tabular}[c]{@{}c@{}} Max $N_b$\\ \end{tabular} & \begin{tabular}[c]{@{}c@{}}Max charges \\per bunch ($|e|$) \end{tabular} \\ \hline \hline
He & 1 & \textcolor{blue}{1.2E+12}, \textcolor{BrickRed}{1.5E+11}                                                        & \underline{\textbf{1.45E+11}}                                                              & \textcolor{blue}{5.1E+10}, \textcolor{BrickRed}{5.1E+10}                                                              & 9.6E+10                                                                  & \textcolor{blue}{1.8E+10}, \textcolor{BrickRed}{1.8E+10}                                                              & \underline{\textbf{1.1E+10}}                                                                    & \textcolor{blue}{7.5E+9}, \textcolor{BrickRed}{7.5E+9}                                                          &      \textcolor{BrickRed}{1.5E+10}                                                     \\
O  & 1 & \textcolor{blue}{1.7E+11}, \textcolor{BrickRed}{\underline{\textbf{2.1E+10}}}                                                        & 3.6E+10                                                             & \textcolor{blue}{1.3E+10}, \textcolor{BrickRed}{7.4E+9}                                                              & 2.4E+10                                                                    & \textcolor{blue}{4.3E+9}, \textcolor{BrickRed}{2.5E+9}                                                              & 2.8E+9                                                                    & \textcolor{blue}{1.9E+9}, \textcolor{BrickRed}{1.7E+9}                                                          &        \textcolor{BrickRed}{1.3E+10}                                                     \\
Mg & 1 & \textcolor{blue}{2.1E+10}, \textcolor{BrickRed}{\underline{\textbf{2.7E+9}}}                                                        & 1.8E+10                                                              & \textcolor{blue}{6.3E+9}, \textcolor{BrickRed}{9.5E+8}                                                              & 1.6E+10                                                                    & \textcolor{blue}{2.4E+9}, \textcolor{BrickRed}{3.6E+8}                                                              & 3.0E+9                                                                    & \textcolor{blue}{1.6E+9}, \textcolor{BrickRed}{2.4E+8}                                                          &    \textcolor{BrickRed}{2.9E+9}                                                       \\
Ar & 1 & \textcolor{blue}{4.1E+10}, \textcolor{BrickRed}{\underline{\textbf{5.1E+9}}}                                                        & 1.2E+10                                                              & \textcolor{blue}{4.2E+9}, \textcolor{BrickRed}{1.8E+9}                                                              & 9.7E+9                                                                   & \textcolor{blue}{1.7E+9}, \textcolor{BrickRed}{7.4E+8}                                                              & 1.9E+9                                                                    & \textcolor{blue}{1.2E+9}, \textcolor{BrickRed}{4.9E+8}                                                          &       \textcolor{BrickRed}{8.9E+9}                                                     \\
Ca & 4 & \textcolor{blue}{1.1E+10}, \textcolor{BrickRed}{5.5E+9}                                                        & \underline{\textbf{5.0E+9}}                                                              & \textcolor{blue}{1.8E+9}, \textcolor{BrickRed}{1.8E+9}                                                              & 1.1E+10                                                                   & \textcolor{blue}{7.6E+8}, \textcolor{BrickRed}{7.6E+8}                                                              & 5.5E+9                                                                    & \textcolor{blue}{5.1E+8}, \textcolor{BrickRed}{5.1E+8}                                                          &             \textcolor{BrickRed}{1.0E+10}                                               \\
Kr & 3 & \textcolor{blue}{1.4E+10}, \textcolor{BrickRed}{\underline{\textbf{5.1E+9}}}                                                        & 6.4E+9                                                              & \textcolor{blue}{2.3E+9}, \textcolor{BrickRed}{1.8E+9}                                                              & 4.5E+9                                                                   & \textcolor{blue}{9.6E+8}, \textcolor{BrickRed}{7.6E+8}                                                              & 8.0E+8                                                                    & \textcolor{blue}{5.4E+8}, \textcolor{BrickRed}{5.1E+8}                                                          &    \textcolor{BrickRed}{1.8E+10}                                                         \\
In & 7 & \textcolor{blue}{5.1E+9}, \textcolor{BrickRed}{4.4E+9}                                                        & \underline{\textbf{3.0E+9}}                                                              & \textcolor{blue}{1.1E+9}, \textcolor{BrickRed}{1.1E+9}                                                              & 3.5E+9                                                                   & \textcolor{blue}{4.6E+8}, \textcolor{BrickRed}{4.6E+8}                                                              & 1.2E+9                                                                    & \textcolor{blue}{3.1E+8}, \textcolor{BrickRed}{3.1E+8}                                                          &   \textcolor{BrickRed}{1.5E+10}                                                         \\
Xe & 6 & \textcolor{blue}{5.8E+9}, \textcolor{BrickRed}{4.3E+9}                                                        & \underline{\textbf{3.1E+9}}                                                              & \textcolor{blue}{1.1E+9}, \textcolor{BrickRed}{1.1E+9}                                                              & 3.1E+9                                                                  & \textcolor{blue}{4.6E+8}, \textcolor{BrickRed}{4.6E+8}                                                              & 8.8E+8                                                                    & \textcolor{blue}{3.1E+8}, \textcolor{BrickRed}{3.1E+8}                                                          &         \textcolor{BrickRed}{1.7E+10}                                                     \\
Pb & 8 & 4.2E+9                                                        & \underline{\textbf{2.6E+9}}                                                              & 9.8E+8                                                              & 9.8E+8                                                                    & 3.9E+8                                                              & \underline{\textbf{3.9E+8}}                                                                    & 2.6E+8                                                          & \textcolor{BrickRed}{2.1E+10} \\  \hline                                 
\end{tabular}
\end{adjustbox}
\end{table*}

The predictions for the baseline production scenario, involving PS splitting and PS-SPS stripping as presently done for Pb, are summarized in Table~\ref{tab:results_baseline}. The numerical algorithm follows the computational flowchart in Fig.~\ref{fig:flowchart}, with ion species input from Table~\ref{tab:species_general}, injector parameters from Table~\ref{tab:injector_parameters} and the number of LEIR injections from Table~\ref{tab:cooling_time_ion_future_species}, following the operationally achieved production of Pb in 2024. The maximum injected bunch intensity $N_b$ in each machine at injection is calculated, both with and without considering the LEIR electron cooling limit. The global bottleneck in the ion injector chain is marked in bold and is underscored. SC limits the intensity in LEIR for He, Ca, In, Xe and Pb, and also in the SPS for He and Pb. On the other hand, O, Mg, Ar and Kr are limited by the LEIR electron cooling times. It should be noted that the short 2024 Mg test run~\cite{ipp_mg7_test_slupecki}, with low intensities in LEIR, confirmed that the given length of the injection plateau is insufficient to inject and cool more than one Linac3 pulse, as predicted in Table~\ref{tab:cooling_time_ion_future_species}.

In this baseline production scheme, Pb reaches the highest number of injected charges per bunch into LHC.  

\subsection{Potential ion production scheme improvements}
\label{sec:Improvements}

The present Pb ion production scheme might be further optimised for new ion species in the injectors to improve the injected LHC bunch intensity. Some of the most realistic technical proposals include:

\begin{itemize}
    \item Omitting PS bunch splitting
    \item Adding a new stripping stage between LEIR and the PS
    \item Optimising the charge-to-mass ratio by varying charge states after stripping in Linac3
    \item Exploring different isotopes to optimise the charge-to-mass ratio
    \item Moving to 25 ns bunch spacing in the LHC 
\end{itemize}

\begin{centering}
\begin{table}[b]
\centering
\caption{Theoretical LEIR-PS stripper foil efficiencies for a given LEIR charge state $q_{\textrm{LEIR}}$ ($* =$ interpolated value).}
\label{table:stripping_efficiencies}
\begin{adjustbox}{width=.6\columnwidth}
\begin{tabular}{rrrrr} \hline
Ion & $Z$ & $A$ & $q_{\textrm{LEIR}}$ & LEIR-PS \\ 
    &     &     &    &  stripping \\
    & & & & efficiency \\   \hline 
He  & 2   & 4  & 1 &90\%          \\  
O   & 8   & 16  & 4 &90\%         \\
Mg$^*$ &  12  & 24 & 7 &90\% \\
Ar  & 18  & 40  & 11 &90\%       \\  
Ca  & 20  & 40  & 17 &   90\%        \\  
Kr  & 36  & 86  & 22  &  87\%         \\  
In  & 49  & 115 &  37 & 75\%     \\  
Xe  & 54  & 129 &  39  &50\%       \\  
Pb  & 82  & 208 &  54  & 50\%      \\ \hline
\end{tabular}
\end{adjustbox}
\end{table}
\end{centering}

Omitting the PS bunch splitting allows for higher extracted bunch intensity, but carries the risk of hitting the SC limit at SPS injection resulting in more losses and emittance blow-up. As an alternative, three bunches extracted from LEIR into PS without splitting could be considered an intermediate option.

Inserting a new stripping foil between LEIR and the PS instead of the present stripping stage between the PS and SPS offers the advantage of higher incoming ion charge states for the PS~\cite{Kroeger_gamma_factory_alternative_stripper_foil}. The higher charge state would allow for higher extracted beam momenta from the PS for a given $B \rho$ (cf.~Eq.~\eqref{eq:Brho}) and hence a higher tolerated intensity at SPS injection when considering SC and IBS. Results from studies on the Pb stripper foil efficiencies along the CERN accelerator complex exist~\cite{kroger2022_gamma_factory_stripping_charge}. In a similar fashion, the theoretical LEIR-PS stripping foil efficiencies in Table~\ref{table:stripping_efficiencies} for each considered ion have been computed using the GLOBAL simulation code, which was developed at GSI based on studies of relativistic heavy ions through matter~\cite{scheidenberger1998_heavy_ion_penetration}. This LEIR-PS stripping scenario may increase SC effects at PS injection before the energy is ramped. The stripping to bare nuclei is also likely to increase the spread in angle and energy but this can be mitigated with appropriate optics as previously done in the PS-SPS transfer line. Further experimental studies are called for to determine the possible impact of this new stripping scenario on the beam quality. Lower magnetic fields at PS injection will be required for some ion species, that must be compatible with the lower limit on the PS dipole B-field. Such beams may also need to be captured and accelerated in the PS with a different radio-frequency $f_{\textrm{RF}}$  and/or harmonic number $h$, which have to be synchronised with LEIR. 

Charge state and isotope optimisation have the potential to improve transmitted bunch intensities across the ion injector chain, and will be discussed in detail in Sec.~\ref{sec:charge_state_and_isotope_scan}. 

As an additional scenario, we consider reducing the bunch spacing at PS extraction to 50 ns, which together with SPS slip stacking results in 25 ns bunch spacing in the LHC. An upgraded PS RF system would be required to produce four 50 ns bunches at PS extraction. Two potential RF upgrade schemes are considered: a) narrow-band, tuneable cavities in the 10-20 MHz range, and b) two new wideband Finemet cavities in the range 9.5-20.5 MHz~\cite{ps_rf_upgrades}. In this scenario, we assume similar bunch intensities as in the baseline scenario. Alternatively, two 50 ns bunches from PS could potentially be extracted without any hardware upgrades, but this would lead to an overly long injection process in the LHC due to the many injections needed with very short trains.

Another alternative to improve ion beam production is a higher LEIR extraction energy, which would also require hardware upgrades with new power converters and might lead to longer LHC filling times (in case longer LEIR cycles would be needed). However,
the main concern with this option is related to the radiation hazard it poses. As the LEIR machine does not have a shielded roof, projectile kinetic energies $E_{\textrm{kin}}$ above 90 MeV/u necessitate further risk assessments and additional personnel protection barriers. The present value of $E_{\textrm{kin}}$ for Pb$^{54+}$ at LEIR extraction is about 72 MeV/u, but many of the ion species in Table~\ref{tab:species_general} have an energy closer to this upper limit of 90 MeV/u, leaving little room for improvement without investing in additional radiation protection measures. 

Based on these considerations, we devise \textbf{five operational scenarios}: 
\begin{itemize}
    \item Scenario 1: the default Pb production scenario with PS bunch splitting and PS-SPS stripping
    \item Scenario 2: no PS splitting with PS-SPS stripping 
    \item Scenario 3: PS splitting with a change in the stripper foil location from PS-SPS to LEIR-PS 
    \item Scenario 4: no PS splitting with LEIR-PS stripping, 
    \item Scenario 5: 25~ns bunch spacing in the LHC with similar bunch intensities as in Scenario 1.
\end{itemize}

\begin{figure}[t!]
  \centering
  \includegraphics[width=.96\columnwidth]{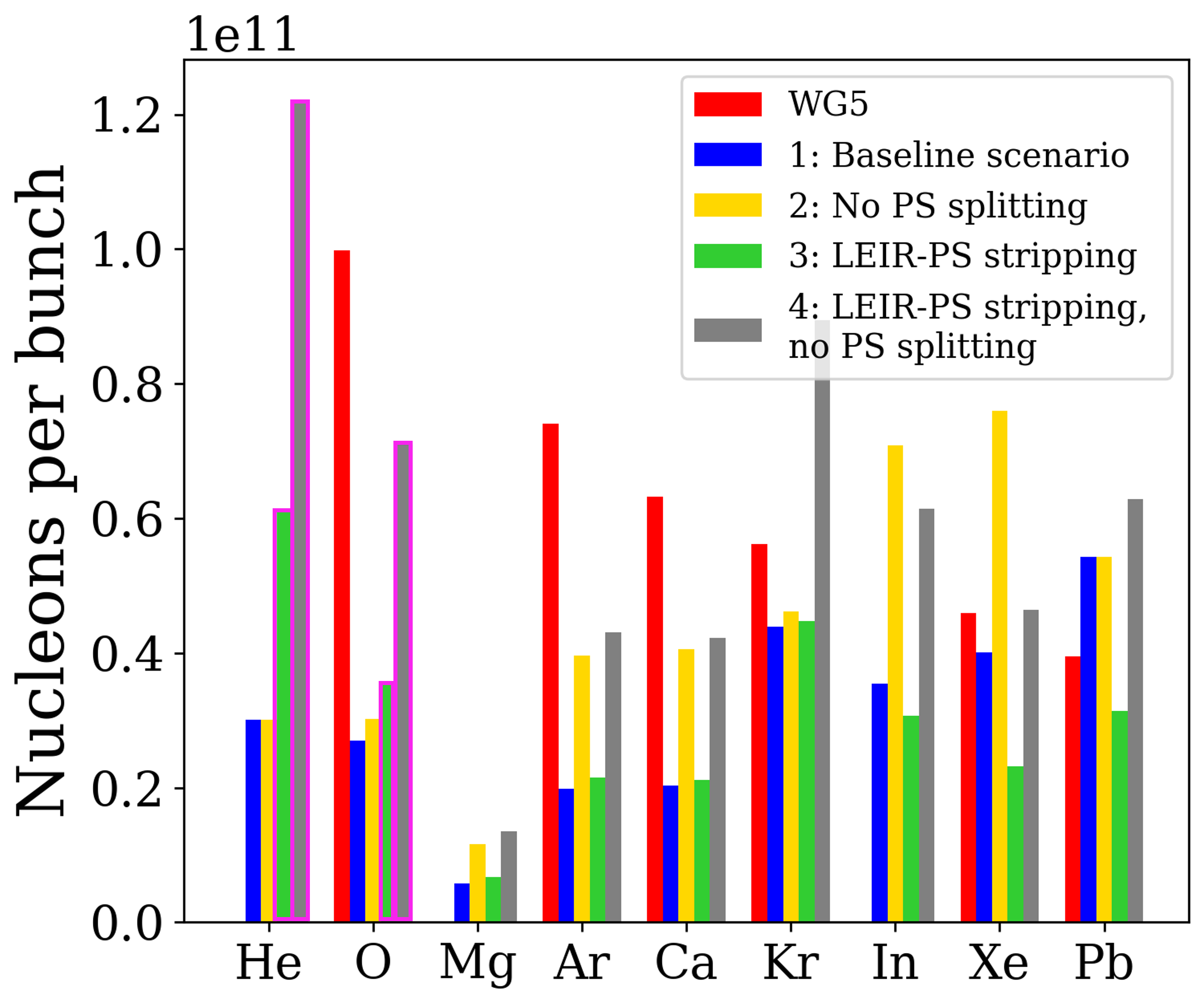}
  \caption{Estimated nucleons per bunch for the different scenarios, including the WG5 estimates~\cite{YR_WG5_2018}. He and O bars with pink frames require a $B_{\textrm{inj}}$ in the PS that presently would be too low.}
\label{fig:all_scenarios}
\end{figure}

The achievable nucleons per bunch, a way to normalise bunch intensity between different ion species, injected in the LHC for the first 4 scenarios are shown in Fig.~\ref{fig:all_scenarios}. The estimated values are all typically lower than the WG5 scenarios~\cite{YR_WG5_2018}, in particular for the lighter ions O, Ar and Ca. However, He, Mg and In were not considered in their study. Only one injection of He, O, Mg and Ar into LEIR would be tolerated with the present electron cooler configuration and LHC filling scheme, causing a large discrepancy compared to the WG5 results.

Stripping between LEIR-PS poses a problem for He and O ions as the dipole magnetic field $B$ at PS injection falls below the minimum threshold $B_{\textrm{min}} = 0.0383$~T in scenarios 3 and 4. These bars are marked in magenta, and require further technical studies on $B_{\textrm{inj}}$ in the PS. For all other ions, the required $B_{\textrm{inj}}$  at PS injection is safely above today's lower limit. 

\subsection{Comparison with 2025 oxygen pilot run data}
\label{sec:oxygen_2025_data}

\begin{table}[b!]
\begin{center}
\caption{Preliminary comparison between: a) typically measured intensities for the single-bunch O$^{4+}$ pilot beam (early July 2025), and b) the most similar production scenario (2: no bunch splitting) from the Injector Model. At LEIR injection (*), the value indicates accumulated intensity before capture. The \textcolor{red}{red} values at SPS injection indicate the space charge limit, versus maximum injectable in black.}
\label{table:oxygen_2025_vs_model_data}
\begin{adjustbox}{width=\columnwidth}
\begin{tabular}{c|ccccc}
{[}ions/bunch{]} & \textbf{LEIR inj.$^{*}$} & \textbf{LEIR extr.} & \textbf{PS inj.} & \textbf{PS extr.} & \textbf{SPS inj.} \\ \hline
Measured & 1.5E+10             & 5.5E+09              & 4.8E+09        & 3.7E+09         & 3.7E+09         \\ \hline
\begin{tabular}[l]{@{}c@{}}Inj. Model \\ (Scenario 2)\end{tabular}     & 2.10E+10             & 8.00E+09              & 7.40E+09        & 5.4E+09         & \begin{tabular}[l]{@{}c@{}}5.0E+09\\ \textcolor{red}{2.8E+09}\end{tabular} \\ \hline    
\end{tabular}
\end{adjustbox}
\end{center}
\end{table}

The commissioning of a single-bunch O$^{4+}$ pilot beam performed in June and July 2025 provides a preliminary check of the Injector Model predictions. Despite using a different beam type and production scheme, some first comparisons can be made. The LEIR beam variant used for this pilot beam captures two bunches with RF harmonic $h=2$, but only transfers one of the bunches to the PS. In the PS, the O$^{4+}$ beam did not undergo bunch splitting and was immediately ramped and then stripped of all remaining electrons during beam transfer to the SPS. In the SPS, a four bunch injection scheme lasting 12~s was deployed, compared to the 14 train injection scheme over 48 s used for \pb. The closest comparable Injector Model scenario is therefore Scenario 2 with no PS bunch splitting. A comparison of typical measured values from July 6$^{\text{th}}$ 2025 (during one of the successful LHC fills) compared to the model values are shown in Table~\ref{table:oxygen_2025_vs_model_data}, obtained with the measured 90 $\mu$A Linac3 current from Table~\ref{tab:species_general}. Despite the fact that the Injector Model assumes a higher intensity at LEIR injection, due to additional variation in injection efficiency observed in the measurements, the total transmission values (extraction/injection ratio) in both LEIR and in the PS model agree quite well with the measurements: 38\% compared to 37\% for LEIR, and 73\% compared to 77\% for the PS. Fewer PS-SPS transfer losses were measured than were assumed in the model. 
The typically measured bunch intensity at SPS injection is in line with the model interval between the maximum possible total bunch intensity of 5.0E+09 and the tolerable space charge limit of 2.8E+09. In the measurements, 
a typical loss rate of 1.5\%/s was observed for O$^{8+}$ beams, compared to around 0.8\%/s for the Pb ion reference case. However, this was for a tune working point that was giving better beam brightness (intensity/emittance) on the short cycle used for the LHC pilot run. A test with the optimal working point for long cycles (similar to the one used for Pb) using the same intensity as in Table~\ref{table:oxygen_2025_vs_model_data}, showed loss rates on the order of 0.5\%/s. This improved transmission compared to the Pb beam is expected from the weaker IBS effects for the lighter ions. 
Overall, these preliminary measurements do not reveal any fundamental gap in the model assumptions. 

\subsection{Charge state and isotope scan} \label{sec:charge_state_and_isotope_scan}

Another possibility to obtain higher LHC bunch intensities is to vary the outgoing Linac3 charge state or isotope. A higher LEIR charge state enables acceleration to higher energies in LEIR and PS, and therefore faster electron cooling according to Eq.~\eqref{eq:ecooling_scaling} Therefore for more injections into LEIR could be allowed than listed in Table~\ref{tab:cooling_time_ion_future_species}. In considering charge state scans, it is vital to take into account previous source experiments~\cite{Scrivens_gamma_factory_source_status, detlef_Kr_source, detlef_Mg_source} that have demonstrated that some lighter ion species do not need stripping after Linac3, with a $q/A$ ratio sufficiently high to be directly injected into LEIR. These ions include He, O, Mg and Ar, while Kr was also successfully injected into LEIR without stripping after re-optimisation of the source.

\begin{figure}[t]
  \centering
  \includegraphics[width=.96\columnwidth]{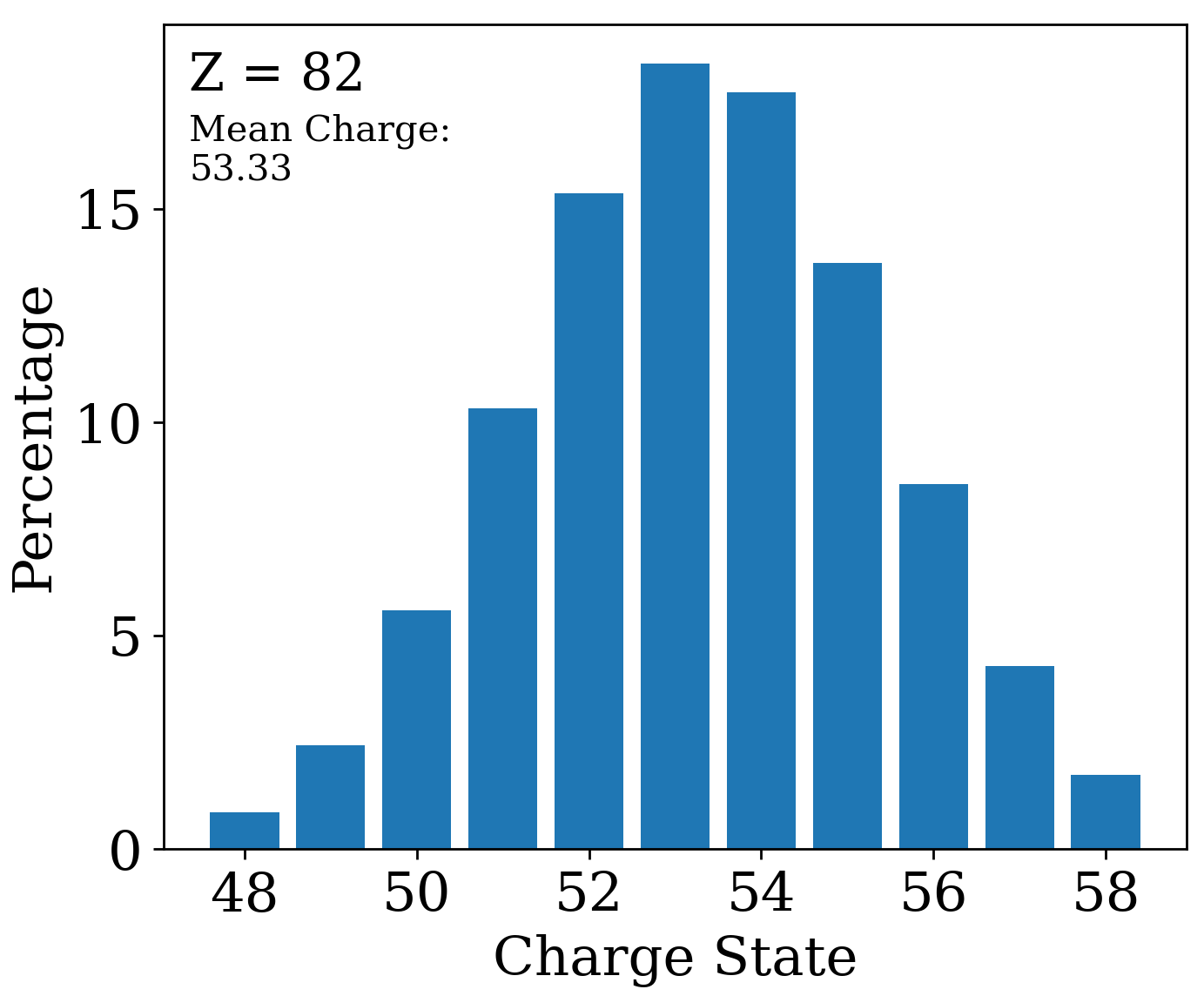}
  \caption{Pb charge state distribution after carbon foil stripping after Linac3 at 4.2 MeV/u, estimated with Baron's formula.}
\label{fig:baron_charge_state_scan_Pb}
\end{figure}

The ions Ca, In, Xe and Pb do require stripping after Linac3, as does Kr$^{29+}$~\cite{bruce_2021_Preliminary_LHC_light_ion_scenarios}. In order to estimate the relative charge state distribution after stripping at the exit of Linac3 and target meaningful stripping efficiencies, we use the established Baron's formula~\cite{baron}, recently implemented numerically in the Python package \verb|baroncs|~\cite{olsen_baroncs_github_repo}. Figure~\ref{fig:baron_charge_state_scan_Pb} shows the resulting charge state distribution of Pb upon exit from Linac3 as an illustrative example. The most abundant charge state, Pb$^{53+}$, was excluded for operation because of its high recombination rates~\cite{leir_ecooling_1999experimental}, which we do not expect to be a problem in this context as no charge state was found to be isoelectronic with Pb$^{53+}$. 
The second-most abundant charge state, Pb$^{54+}$, is currently used for today's standard ion physics operation. All other most abundant charge states Ca$^{17+}$, In$^{37+}$ and Xe$^{39+}$ predicted by \verb|baroncs| agree with the default charge states listed in Table~\ref{tab:species_general}.

\begin{figure}[b]
  \centering
  \includegraphics[width=\columnwidth]{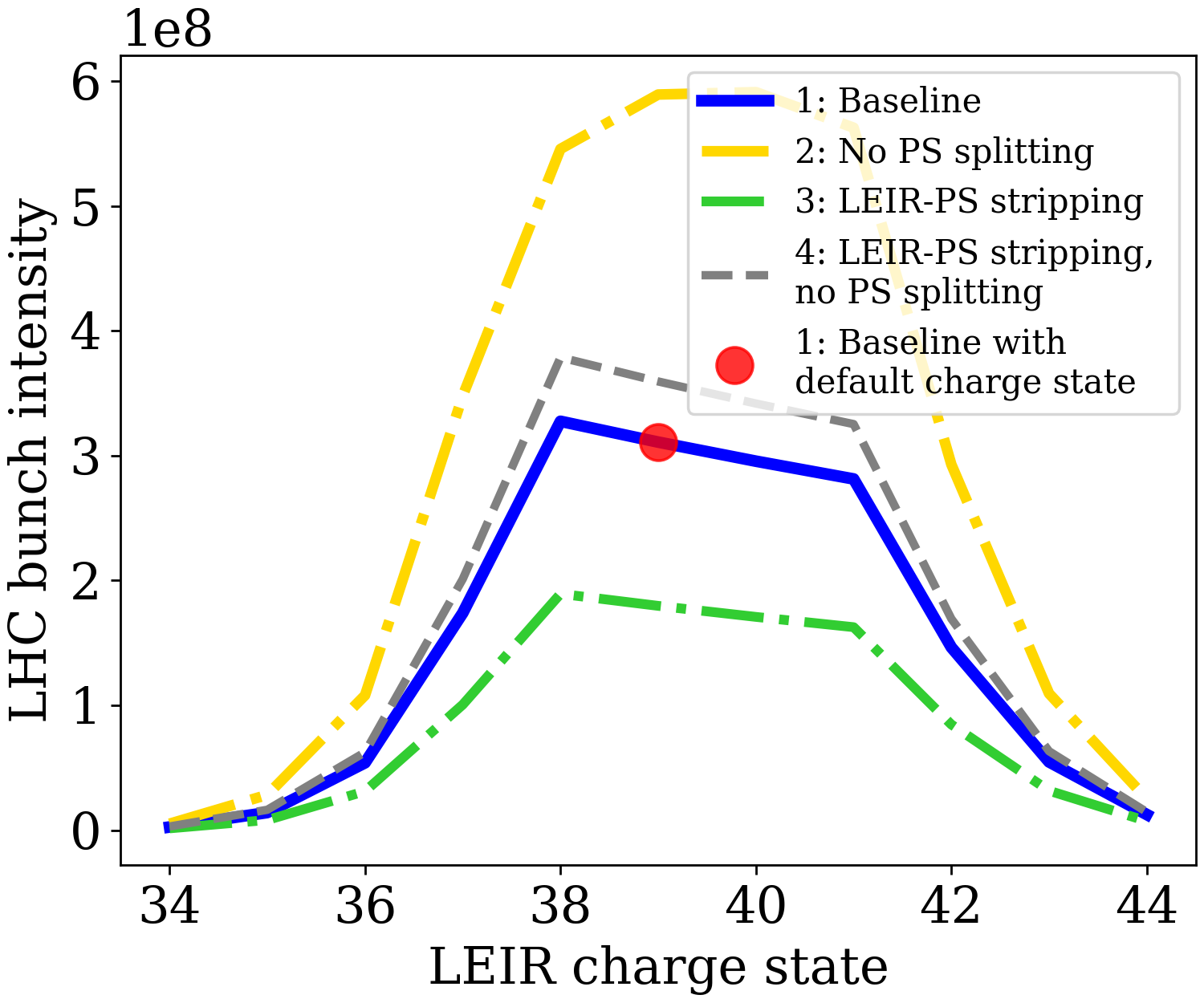}
  \caption{Xe charge state scan: injected ions per bunch into LHC for different LEIR charge states.}
\label{fig:Xe_charge_state_scan}
\end{figure}

An illustrative example of the importance of optimizing the charge state in LEIR is demonstrated for Xe in Fig.~\ref{fig:Xe_charge_state_scan}. The colored curves representing the four scenarios as computed by the Injector Model contain numerous non-linear superpositions of SC limits, different numbers of LEIR injections (due to different cooling performance), and Linac3-LEIR stripping efficiencies from Baron's formula. For the baseline scenario, Xe$^{38+}$ may offer small gains in injected LHC intensity, but Xe$^{39+}$ remains the superior charge state for Scenario 2 with no PS bunch splitting. 

\begin{figure}[b]
  \centering
  \includegraphics[width=\columnwidth]{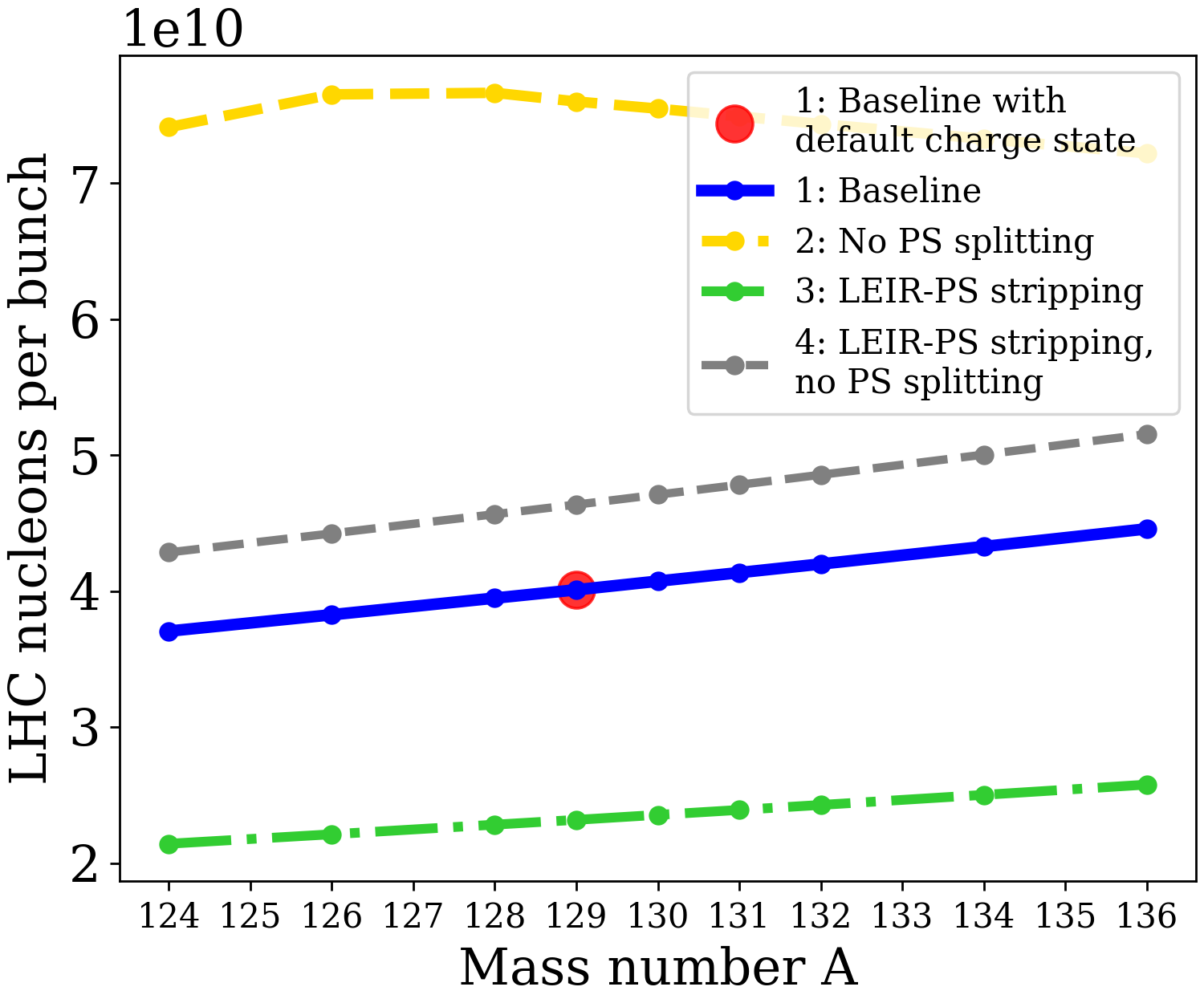}
  \caption{Xe isotope scan: injected nucleons per bunch into LHC for different stable isotopes (with dotted markers).}
\label{fig:Xe_isotope_scan}
\end{figure}

In the second step, we scan all stable isotopes for each ion species and find that some gain in injected nucleons per bunch into the LHC $\mathcal{N}_{b,0}$ can be achieved by changing isotope, depending on the scenario. Figure~\ref{fig:Xe_isotope_scan} shows an example isotope scan for Xe, where changing the default isotope $^{129}$Xe to $^{128}$Xe increases the injected nucleons per bunch by 3\% for Scenario 2 (no PS bunch splitting). In terms of injected bunch intensity, the isotope $^{3}$He delivers 33\% more than $^{4}$He, $^{78}$Kr 10\% more than $^{86}$Kr and $^{204}$Pb 2\% more than $^{208}$Pb. The extent to which these isotopes can be acquired at reasonable costs and their possible implementation in the source has not yet been investigated. For instance, the natural abundance of $^{3}$He is only 0.0002\%, $^{78}$Kr only 0.36\%, $^{128}$Xe only 1.9\%, and $^{204}$Pb only 1.4\%~\cite{kondev2021_nuclear_database}. The price and availability typically depend not only on abundance but also on how easily they separate. A full compilation of the charge scan and isotope results can be found in the Appendix.

The condensed results of the charge state and isotope scan are summarised in Table~\ref{table:charge_state_and_isotope_scan}, which shows the best LEIR charge state $q_{\mathrm{LEIR}}$ and mass number $A$ in Scenario 2 (no PS bunch splitting). The relative increase for the injected bunch intensity into the LHC with the new charge state or isotope compared to the default choices from Table~\ref{tab:species_general} are also computed. Changing charge states can give considerable gains, and appears to be more easily implementable and less costly than the purchase of exotic isotopes. Note that Pb$^{56+}$ achieves 12\% more intensity than Pb$^{54+}$ for scenario 2 thanks to a higher space charge limit at SPS injection (the global bottleneck for most Pb charge states) and despite its lower stripping efficiency as seen in Fig.~\ref{fig:baron_charge_state_scan_Pb}.

\begin{table}[t]
\caption{Ratio between injected nucleons per bunch into the LHC $\mathcal{N}_{b,0}$ for default values in Scenario 2 (no PS splitting), versus using the calculated a) best LEIR charge state $q_{\mathrm{LEIR}}$ and b) the best isotope with mass number $A$.}
\centering
\begin{adjustbox}{width=.9\columnwidth}
\centering
\begin{tabular}{l|lll|lll} \hline
Ion & $q_{\mathrm{LEIR}}^{{\mathrm{best}}}$ & $q_{\mathrm{LEIR}}^{{\mathrm{0}}}$ & $N_{b,q}$/$N_{b, 0}$ & $A_{{\mathrm{best}}}$ & $A_0$ & $\mathcal{N}_{b, A}$/$\mathcal{N}_{b, 0}$ \\ \hline \hline
He  & 1 & 1  & 1.0 & 3 & 4 & 1.33 \\
O   & 5 & 4 & 1.35 & 16 & 16 & 1.0 \\
Mg  & 7 & 7 & 1.0 &  24 & 24 & 1.0 \\
Ar  & 11 & 11 & 1.0 & 36 & 40 & 1.33 \\
Ca  & 17 & 17 & 1.0 & 43 & 40 & 1.16 \\
Kr  & 28 & 22 & 1.46 & 78 & 86 & 1.10\\
In  & 36 & 37 & 1.06 & 115 & 115 & 1.0\\
Xe  & 39 & 39 & 1.0 & 128 & 129 & 1.01 \\
Pb  & 56 & 54 & 1.12 & 204 & 208 & 1.02 \\ \hline  
\end{tabular}
\end{adjustbox}
\label{table:charge_state_and_isotope_scan}
\end{table}

The charge state scan also reveals whether stripping or direct source production is the most beneficial for Kr. Despite a higher deliverable Linac3 current of 40 $\mu$A confirmed for Kr$^{22+}$ as seen in Table~\ref{tab:species_general}, the previously assumed charge state Kr$^{29+}$ (with 25 $\mu$A Linac3 intensity assumed in the absence of experimental data) has two main advantages: the higher charge state allows for two extra injections into LEIR (at least five instead of three) and leads to a higher SPS SC limit, as the kinetic energy per nucleon at PS extraction will be higher. For this reason the injected LHC bunch intensity for Kr$^{29+}$ is estimated to be 39\% more and for Kr$^{28+}$ 46\% more in comparison with Kr$^{22+}$ . The full parameter scan results for various charge state and isotopes are found in the Appendix. 

Operation will not be possible for all charge states~\cite{Scrivens_gamma_factory_source_status}. Key issues for some charge states are the potentially reduced stripping efficiency and hardware limitations such as the fact that the Linac3-LEIR transfer line cannot handle ion charge states with a ratio $q/A < 0.25$. In addition, charge over mass ratios $q/A \approx 0.3$ lead to $E_{\textrm{kin}}$ above 90 MeV/u at LEIR extraction and hence additional radiation constraints. For example, O$^{5+}$ reaches kinetic energies of 103 MeV/u at LEIR extraction, whereas higher charge states are even more energetic. Without major radiation protection hardware upgrades in LEIR, O$^{6+}$ or any higher oxygen charge states are excluded, despite their attractiveness. Radiation limits from neutron production in Linac3 can also be an issue for some lighter ions. 

We do not exclude future source tuning and optimisation. For instance, He and In currents from~\cite{Scrivens_gamma_factory_source_status} were measured with the previous ECR source, and the current source may deliver higher currents. In the same study, the reduced plasma chamber lifetime observed with Ar lowered the peak performance, meaning that realistic currents for Ar could be higher than in Table~\ref{tab:species_general}. However, we conservatively assume that some charge states of He, O, Ar, Mg and Kr will require hardware upgrades, such as LEIR additional personnel protective barriers, improved vacuum systems and new power converters for the Linac3 transfer line, for which we encourage further operational feasibility studies for particularly interesting charge states.

\section{Studies of LHC luminosity}
\label{sec:LHC}

The achievable luminosity in the LHC can be estimated starting from the studies of achievable bunch intensities from the injectors. The Collider Time Evolution (CTE) software~\cite{bruce21_EPJplus_ionLumi, bruce10prstabCTE} is used to simulate the LHC beam and luminosity evolution for typical physics fills with various ion species. CTE is a particle tracking simulation code that includes physical processes such as luminosity burn-off, IBS and synchrotron radiation damping. The code has been extensively benchmarked with LHC data~\cite{bruce21_EPJplus_ionLumi}. 

We simulate the most promising ion species for three different scenarios of injector and LHC performance, described below. Some assumptions on the LHC parameters and configuration are common to all ion species and all scenarios. Unless explicitly mentioned otherwise, the machine parameters are taken from Table~14 in~\cite{bruce20_HL_ion_report} and are identical to Pb-Pb operation in LHC Runs~3--4. We assume that the machine configuration (optics, RF settings, etc.) is the same for all cases, except for a smaller $\beta^*$ in one scenario, and that the experimental detectors will no longer be limited in terms of event rate~\cite{alice_2022_letter_of_intent}. Furthermore, as observed previously with Pb beams~\cite{bruce21_EPJplus_ionLumi}, we assume additional beam losses corresponding to a non-collisional lifetime of 100~h. The collisional losses are assumed not to be limiting, so no luminosity levelling is used. The interaction cross sections used to simulate the burn-off in the collisions are taken from Table~3 in ~\cite{YR_WG5_2018}. 

In all LHC scenarios and for all ion species, we assume the same transverse geometric emittance at injection from the SPS as obtained operationally with Pb in 2024 ($1.18\times10^{-8}$~m), since the detailed emittance evolution through the whole injector complex cannot yet be simulated in a reliable way. CTE is then used to simulate the beam evolution on the LHC injection plateau, resulting in an emittance blow-up and intensity loss from debunching for each ion and each scenario as a function of the assumed injection time. This, together with an ion-independent transmission of 97\% during LHC acceleration (as observed in 2024), gives the starting conditions in collision. From a second CTE study of the beam evolution during collisions, we then estimate the total achievable luminosity during a typical one-month run as in~\cite{bruce21_EPJplus_ionLumi}, including $T_\mathrm{run}=24$~days of physics data taking and one day for van-der-Meer scans with very small luminosity production. We  first calculate an optimal fill length $T_\mathrm{f,opt}$ (time spent in collision before the beams are dumped, based on the simulated luminosity at the ALICE experiment) for each studied case. The calculation assumes a given turnaround time, i.e. the time needed to get back to collisions after the end of the previous fill, detailed below for each scenario.  From the luminosity produced per fill, the optimal fill length and the turnaround time, the time-averaged instantaneous luminosity $\Lu_\mathrm{avg}$ is calculated. The total integrated luminosity $\Ltot$ for a total running time $T_\mathrm{run} $ is then 
\begin{equation}
 \Ltot \, = \, \Lu_\mathrm{avg} (T_\mathrm{f,opt}) \times T_\mathrm{run} \times \eta,
 \label{eq:Ltot}
\end{equation}
where $\eta$ is the operational efficiency, accounting for downtime and unavailability of the machine, premature fill aborts, occasional longer turnaround times as well as the gradual build-up of performance during the few weeks of the run. 

\begin{figure}[b]
  \centering
  \includegraphics[width=.96\columnwidth]{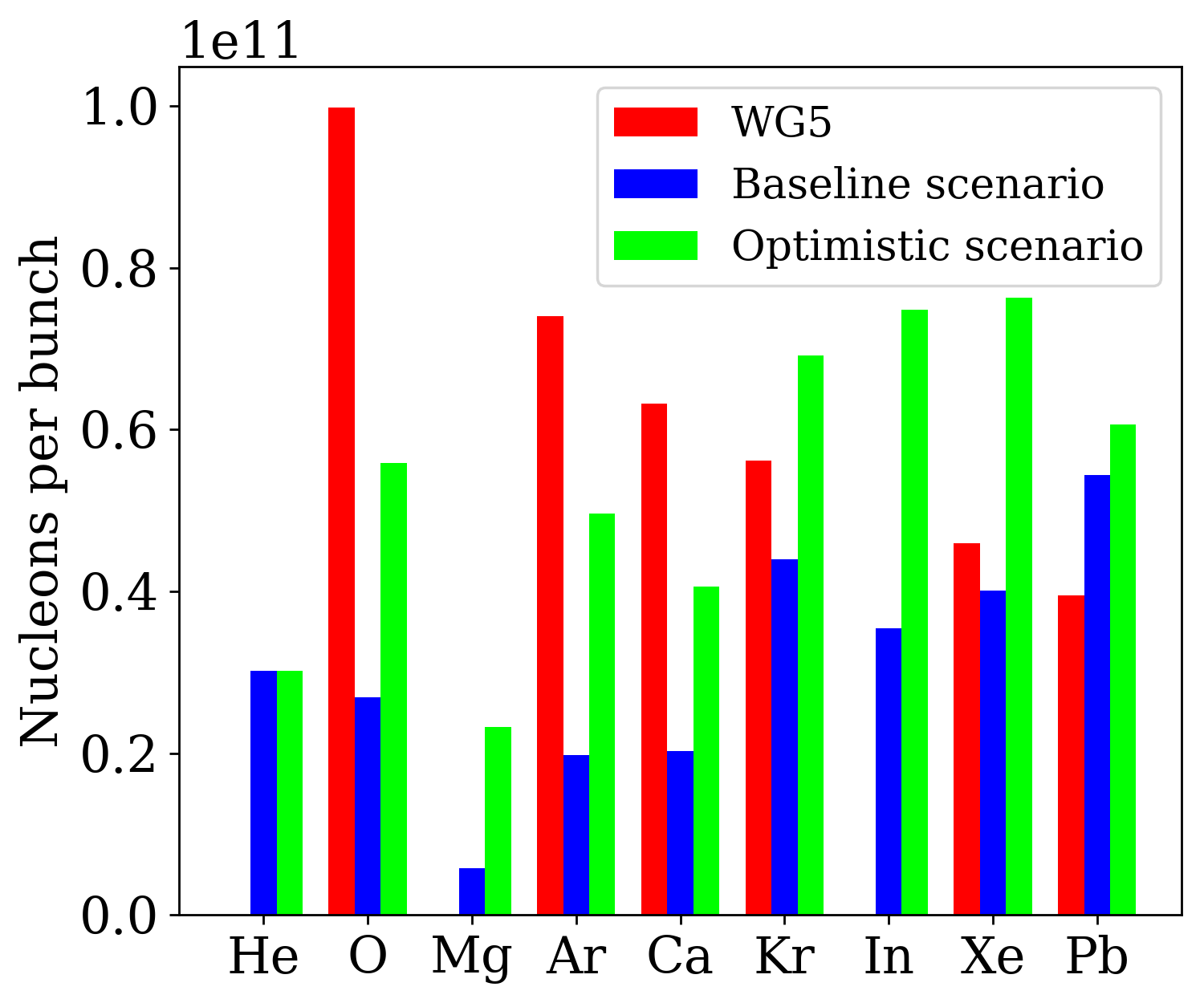}
  \caption{Estimated number of nucleons per bunch injected into LHC from WG5 versus (1) the baseline scenario from Fig.~\ref{fig:all_scenarios} and (2) an ``optimistic" scenario with an improved charge state and an additional LEIR injection, if applicable.}
\label{fig:baseline_vs_optimic_scenario}
\end{figure}

We define \textbf{three different performance scenarios} for the injectors and the LHC itself. Two scenarios, based on achievable injected bunch intensity into the LHC, are presented in Fig.~\ref{fig:baseline_vs_optimic_scenario}. An additional scenario with reduced PS bunch spacing is also tested. The cases are as such:

\begin{itemize}
    \item \textbf{Baseline scenario}: Bunch intensities are calculated using the same production scheme as assumed for future Pb runs and using our best available knowledge today of the injector limitations. This corresponds to the baseline scenario defined in Sec.~\ref{sec:BaselineInt}, also represented by blue bars in Fig.~\ref{fig:baseline_vs_optimic_scenario}. For the LHC, we base the assumptions on what has been achieved with Pb in 2024: the same optics ($\beta^*=50$~cm at ALICE, ATLAS and CMS, with 150~$\mu$rad half crossing angles), and the same filling scheme based on slip-stacking in the SPS (1240~bunches, with 1032 collisions in ALICE, ATLAS, and CMS, and 557 collisions at LHCb). A beam energy of 6.8~$Z$~TeV is assumed, which is the present baseline for Run~4. We assume, as for Pb in the HL-LHC, a projected turnaround time of 3.33~h~\cite{jowett17_cham} and an operational efficiency $\eta=50$\%~\cite{bruce21_EPJplus_ionLumi}.
    
    \item \textbf{Optimistic scenario}: This scenario assumes running with the the best possible charge state in LEIR from Table~\ref{table:charge_state_and_isotope_scan}, and suppressing the PS bunch splitting from Sec.~\ref{sec:Improvements}, exploiting the fact that the SPS is not running at the SC limit in the baseline scenario (for all ion species apart from He, O and Pb). In addition, an extra injection from Linac3 thanks to improved LEIR electron cooling performance is assumed (the maximum is still~8 in total), except for He that has an excessively long cooling time. For the LHC, we assume that in Run~5 some new developments and improvements can be implemented. In particular, 
    we assume that the operational efficiency can be increased to $\eta=65$\%, which was already demonstrated in the 2018 Pb-Pb run, that the beam energy can be increased to the LHC design value of 7~$Z$~TeV, that the external crossing angles can be reduced to 130~$\mu$rad, and that the optics can be squeezed further to give a $\beta^*=40$~cm. Using the methods in~\cite{bruce15_PRSTAB_betaStar, bruce17_NIM_beta40cm} this value seems compatible with the measured aperture in the LHC in the 2024 Pb-Pb run, although further measurements are needed in Run~4 to confirm this. Given the different bunch structure in the SPS without PS splitting, we assume 15\% fewer bunches in the LHC (i.e.~1054~bunches, with 877 collisions in ALICE, ATLAS, and CMS, and 473 collisions at LHCb). In this scenario, more injections are needed into the LHC, leading to an increased turnaround time estimated at 4~h. This leads to increased losses and emittance blow-up on the injection plateau, accounted for each ion species through the individual CTE simulations over the longer LHC injection plateau. A detailed LHC filling scheme is still to be worked out considering the shorter SPS batches used in this scenario. 
    
    \item \textbf{25 ns bunch spacing}: This scenario, based on a possible future upgrade of the PS RF system, optimistically assumes identical bunch intensities and emittance as in the baseline scenario but with more tightly spaced bunches, leading to a filling scheme with 1736 bunches and 1680 collisions at ALICE. The main advantage of this scheme is the larger number of colliding bunches giving higher luminosity, but LHC filling time increases compared to the baseline scenario meaning that the turnaround time would go up to about 4~h. As in the optimistic scenario, this longer injection plateau leads to more losses and higher emittance, something which is simulated individually for each ion species. All other assumptions are identical to the baseline scenario.
\end{itemize}

It should be stressed that the optimistic scenario relies on untested configurations and concepts, and that the 25~ns scenario relies on hardware upgrades in the PS that are not in the present upgrade plan.
Although these improvements have some potential to be implemented, they cannot be taken for granted, while the baseline scenario is based on our current experience of what can realistically be achieved. The achievable luminosity for each ion species may therefore lie somewhere in between these scenarios. 

For all scenarios it is assumed that the LHC can handle the higher intensities and luminosities foreseen. This needs further feasibility studies, noting, in particular, that the collimation of ion beams poses many challenges compared to protons~\cite{bruce14_PRSTAB_sixtr, hermes16_nim}, coming from the nuclear fragmentation of ions entering collimators. The handling of beam losses in general needs careful study, e.g. the energy deposition and radiation caused by collision products. For the present LHC operation with Pb, the secondary beams produced through bound-free pair production and electromagnetic dissociation risk to quench LHC magnets at nominal luminosity unless mitigation measures are applied~\cite{prl07, prstabBFPP09,schaumann20_PRAB_BFPP}. These processes will be much less important for lighter ions, as shown by the cross sections in Table~3 in~\cite{YR_WG5_2018}. However, the power load from luminosity debris lost near the collision points will increase significantly. The beam-induced experimental backgrounds with other ion species will also need to be studied. 

\begin{table*}[ht]
\centering
\caption{Projections of one-month integrated ion-ion luminosities $\int \mathcal{L}_{AA} dt$ and one-month integrated nucleon-nucleon luminosities $\int \mathcal{L}_{NN} dt$ from WG5~\cite{YR_WG5_2018} versus the baseline, optimistic and 25 ns scenarios from this study.}
\begin{adjustbox}{width=0.78\linewidth}
\centering
\begin{tabular}{c|cccc|cccc} \hline
   & \multicolumn{4}{c}{One-month $\int \mathcal{L}_{AA} dt$ [nb$^{-1}$]} & \multicolumn{4}{c}{One-month $\int \mathcal{L}_{NN} dt$ [pb$^{-1}$]} \\ \hline
   & WG5           & Baseline         & Optimistic        & 25 ns & WG5            & Baseline        & Optimistic & 25 ns        \\ \hline \hline
O  & 11700 & 428.1& 2070.4 & 648.6 & 2995.2 & 109.6  & 530.0 & 166.0 \\
Ar & 1080  & 38.0  & 244.7  & 58.3  & 1728.0 & 60.8  & 391.4 & 93.2  \\
Ca & 799   & 48.4  & 199.3  & 73.3  & 1278.4 & 77.5  & 318.9 & 117.2 \\
Kr & 123   & 29.4  & 70.9   & 40.3  & 748.3  & 217.5 & 524.5 & 298.2 \\
In & -     & 10.7   & 36.7   & 14.5  & -    & 141.5 & 485.2 & 192.1 \\
Xe & 28.9  & 9.3   & 26.6   & 12.6  & 480.9  & 154.3 & 443.4 & 209.7 \\
Pb & 4.92  & 3.0   & 3.5    & 3.7   & 212.9  & 131.8 & 153.5 & 160.2 \\ \hline
\end{tabular}
\end{adjustbox}
\label{table:luminosities}
\end{table*}

\begin{figure}[b]
  \centering
  \includegraphics[width=.96\columnwidth]{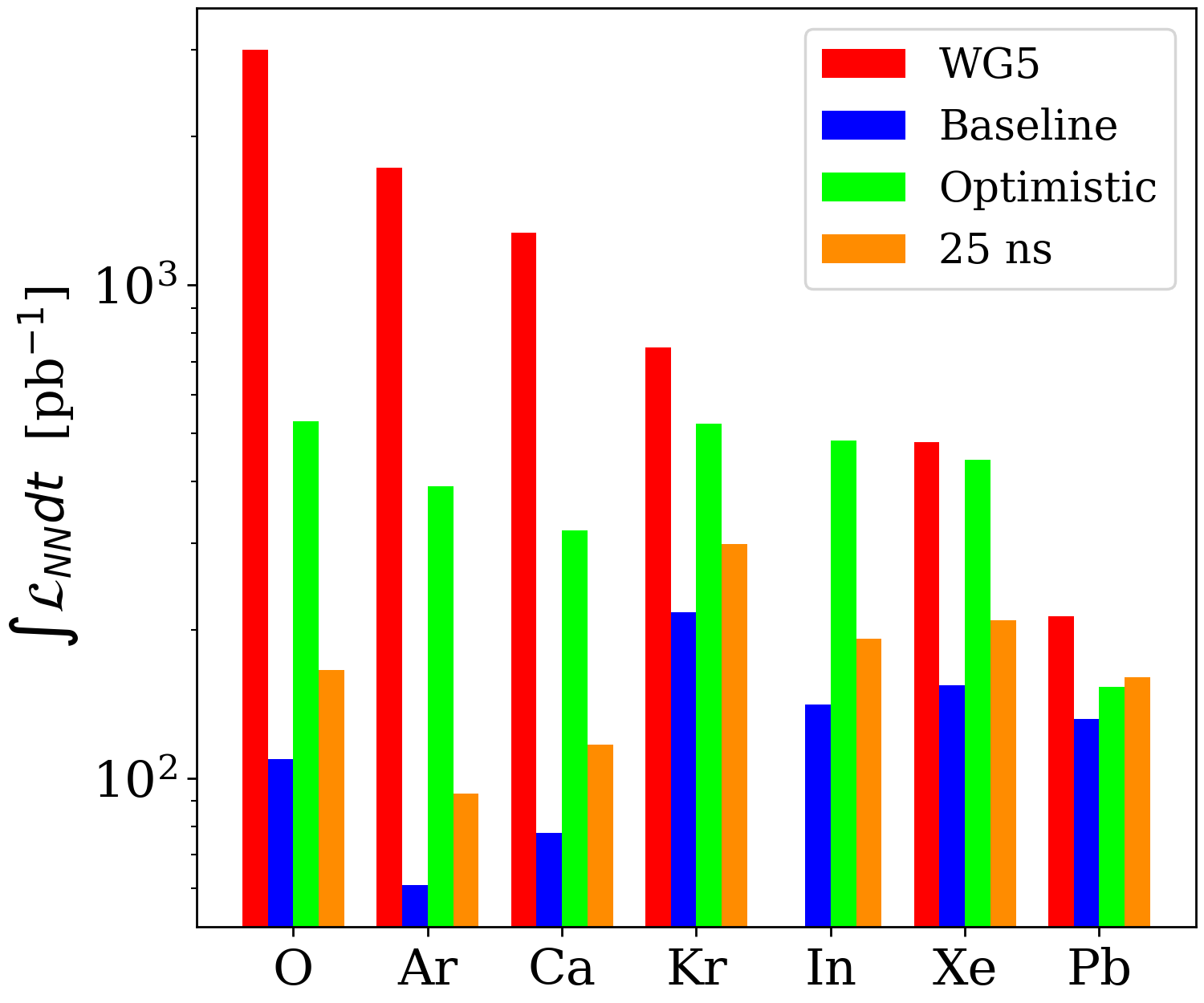}
  \caption{One-month integrated nucleon-nucleon luminosity $\int \mathcal{L}_{NN} dt$ in [pb$^{-1}$] for the various cases.}
\label{fig:L_NN_one_month}
\end{figure}

The resulting projected one-month integrated ion-ion luminosities $\int \mathcal{L}_{AA} dt$ and the nucleon-nucleon luminosities $\int \mathcal{L}_{NN} dt$ are shown in Table~\ref{table:luminosities}, for the ALICE experiment. The other high-luminosity experiments are expected to obtain similar but slightly lower values. The results are graphically displayed in Figs.~\ref{fig:L_AA_one_month} and~\ref{fig:L_NN_one_month}. Peak luminosities, event rates and assumed hadronic cross sections are found in Table~\ref{table:peak_event_rates}. Several observations can be made:

\begin{figure}[b]
  \centering
  \includegraphics[width=.96\columnwidth]{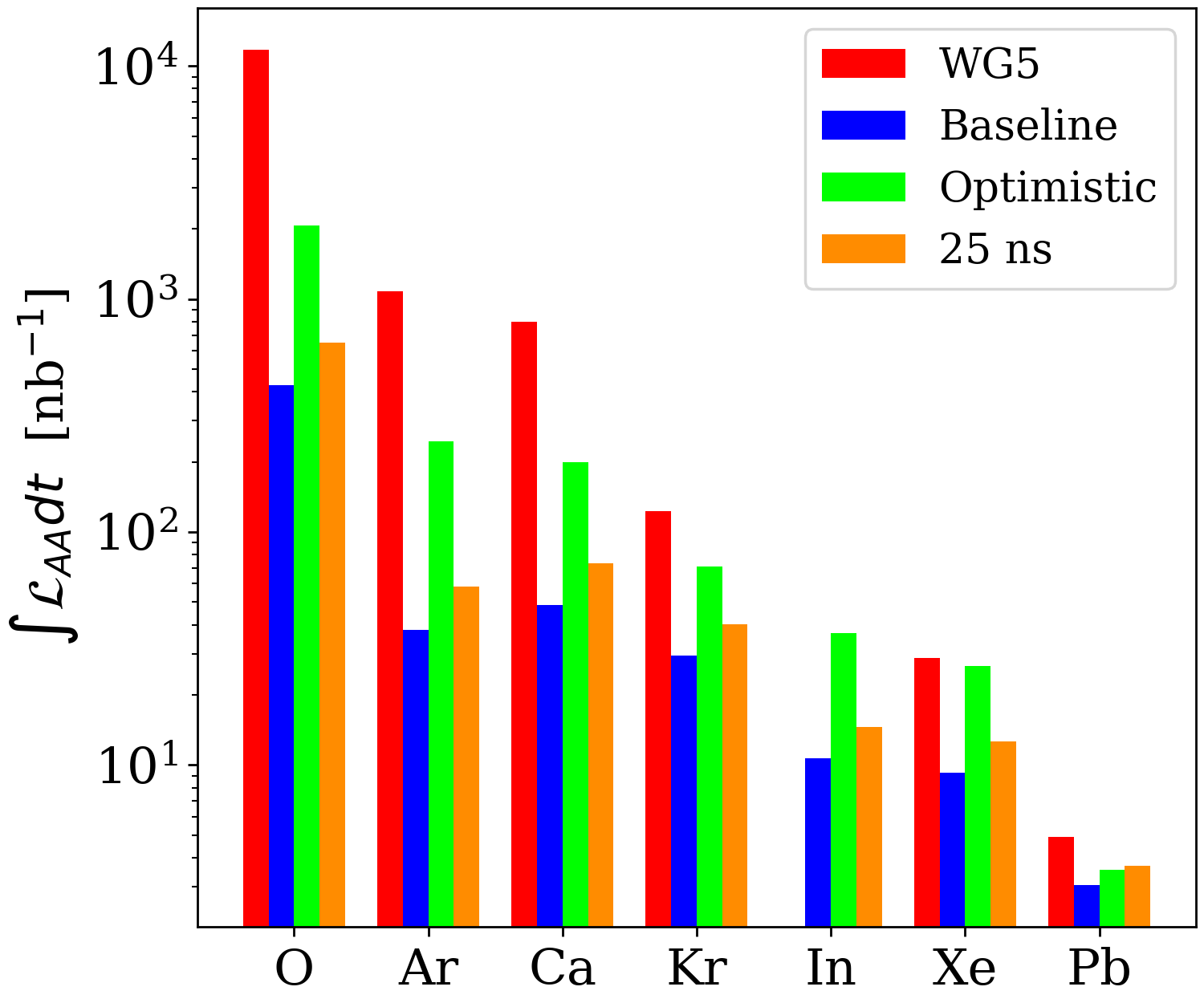}
  \caption{One-month integrated ion-ion luminosity $\int \mathcal{L}_{AA} dt$ in [nb$^{-1}$] for the various cases.}
\label{fig:L_AA_one_month}
\end{figure}

\begin{itemize}
    \item The computed  $\int \mathcal{L}_{AA} dt\approx3$~\nb for Pb in the baseline case in Table~\ref{table:luminosities} is about 20\% higher than previous predictions in Table 6 of~\cite{bruce21_EPJplus_ionLumi}. This comes from our assumed bunch intensity, based on what was achieved in 2024, which significantly surpassed the projected HL-LHC intensity used in~\cite{bruce21_EPJplus_ionLumi}, as well as the absence of luminosity levelling.
    
    \item The estimated $\int \mathcal{L}_{NN} dt$ is lower than reported in WG5~\cite{YR_WG5_2018} for all ions, even in the most optimistic scenario. The discrepancy is largest for lighter ions, 
    with WG5 suggesting a 14-fold gain for oxygen compared to lead, in contrast to our calculations summarised in Table~\ref{table:luminosities} that show that the $\int \mathcal{L}_{NN} dt$ improvement factor for O compared to Pb is 3.5 in the optimistic scenario, and even lower than Pb in the baseline scenario.
\end{itemize}

\begin{table*}[ht]
\centering
\caption{Peak luminosities $\mathcal{L}$ (Ion-ion $AA$ and nucleon-nucleon $NN$) and ion-ion hadronic event rates $N_{AA \mathrm{, had}}$ for baseline, optimistic and 25 ns scenarios, as well as the hadronic cross sections $\sigma_{\mathrm{had}}$.}
\begin{adjustbox}{width=\linewidth}
\centering
\begin{tabular}{c|ccc|ccc|ccc|c} \hline
   & \multicolumn{3}{c}{$\mathcal{L}_{AA}$ [cm$^{-2}$s$^{-1}$]} & \multicolumn{3}{c}{$\mathcal{L}_{NN}$ [cm$^{-2}$s$^{-1}$]} & \multicolumn{3}{c}{$N_{AA \mathrm{, had}}$ [kHz]} & $\sigma_{\mathrm{had}}$ [barn] \\ \hline
   & Baseline            & Optimistic            & 25 ns              & Baseline            & Optimistic            & 25 ns              & Baseline     & Optimistic     & 25 ns    &                \\ \hline \hline
O  & 5.8E+29 & 2.6E+30 & 9.2E+29 & 1.5E+32 & 6.6E+32 & 2.4E+32 & 819.7 & 3639.9 & 1299.35 & 1.41 \\
Ar & 4.8E+28 & 2.9E+29 & 7.7E+28 & 7.7E+31 & 4.6E+32 & 1.2E+32 & 124.7 & 751.0  & 199.1 & 2.60  \\
Ca & 5.2E+28 & 2.0E+29 & 8.2E+28 & 8.3E+31 & 3.2E+32 & 1.3E+32 & 134.0 & 524.9  & 212.1 & 2.60  \\
Kr & 4.6E+28 & 1.0E+29 & 6.6E+28 & 3.4E+32 & 7.5E+32 & 4.9E+32 & 198.0 & 441.6  & 286.1 & 4.33 \\
In & 1.8E+28 & 6.7E+28 & 2.6E+28 & 2.4E+32 & 8.8E+32 & 3.5E+32 & 93.4  & 349.6  & 138.3 & 5.25 \\
Xe & 1.7E+28 & 5.3E+28 & 2.6E+28 & 2.9E+32 & 8.8E+32 & 4.3E+32 & 98.2    & 301.6    & 145.2 & 5.67 \\
Pb & 1.1E+28 & 1.2E+28 & 1.5E+28 & 4.8E+32 & 5.2E+32 & 6.5E+32 &  86.7  & 94.4    & 117.3 & 7.80  \\ \hline
\end{tabular}
\end{adjustbox}
\label{table:peak_event_rates}
\end{table*}

\begin{itemize}
    \item Unlike the WG5 study, our analysis does not identify a single standout ion species that far outperforms the others in terms of one-month integrated nucleon-nucleon luminosity. Instead, similar $\int \mathcal{L}_{NN} dt$ is observed for several species, with 530~pb$^{-1}$ being the highest in the optimistic scenario (for O) and 217~pb$^{-1}$ the highest in the baseline (for Kr).
    
    \item The 25 ns scenario typically falls between the baseline and optimistic scenario, except for Pb where it slightly exceeds the optimistic case. Combining the 25 ns scenario with assumptions for the optimistic scenario could yield even higher intensities, which will be looked at in future studies.
    
    \item The limited number of Linac3 injections imposed by the LEIR electron cooler acts as one of the biggest performance hurdles for lighter ions. For He and O, adding one extra injection could improve performance but the resulting beams may not be sufficiently cooled. Future studies should investigate a) to which extent larger emittances from insufficient cooling could be tolerated, b) whether this is compensated elsewhere downstream in the ion injector chain, and c) the trade-off between longer LEIR cycles for further He and O cooling versus longer LHC filling time.
\end{itemize}

\section{Discussion and Conclusions}
\label{sec:conclusion}

The LHC ion physics programme has so far consisted mainly  of Pb-Pb and p-Pb nuclear collisions. A range of lighter ions species has been studied for the ALICE 3 detector upgrade proposal to further increase bunch intensity and hence the integrated nucleon-nucleon luminosity for Run 5 and beyond. 

In this paper, we present new bunch intensity estimates at LHC injection for different ion species using five main ion beam production scenarios in the CERN ion injector chain (source, Linac3, LEIR, PS, SPS). Realistic performance-limiting effects from space charge, IBS, charge-changing beam-gas interactions and electron cooling are discussed and considered in the calculations. Starting from these bunch intensities, we also present LHC luminosity projections. Using the best knowledge at the time of writing, we investigate several technical proposals to improve injected LHC bunch intensities: no bunch splitting in the PS, alternative LEIR-PS stripping, alternative charge states and isotopes in LEIR, and an RF upgrade in the PS to allow 25~ns bunch spacing. This study provides the first indications of where the ion production scheme could be optimised and tailored for each ion species.

We condense the findings into a baseline, an optimistic, and a 25-ns ion production scenario. The baseline is based on the present well-established Pb ion production scheme, while the optimistic consists of new ion injector concepts (no PS bunch splitting, changed LEIR charge state) and more optimistic LHC parameters, defining two boundaries in between which the real achievable luminosity may lie. We find that many of the lighter ions are limited by the LEIR electron cooler in terms of number of injections from Linac3, effectively acting as the global bottleneck for bunch intensities injected in the LHC. We expect also space charge in LEIR and SPS to act as a limiting factor for some ion species. Overall, both baseline and optimistic nucleon-nucleon luminosity projections are lower than the WG5 estimates, in particular for the lighter ions. The 25-ns scenario delivers nucleon-nucleon luminosity gains of 20--50\% compared to the baseline scenario, and the optimistic scenario gives gains by more than a factor~4 for some ions compared to Pb in the baseline scenario, but both scenarios entail longer turnaround time and hence more beam degradation. These updated luminosity projections provide valuable inputs for the design and discovery potential of the future LHC experiments.  

Future research should focus on experimentally validating the Injector Model with additional ion beam tests. A new additional Linac3 source, connection of ion sources and new beam diagnostic lines would allow for parallel commissioning of several ion species in the same year and gaining operational experience with new ion beams. Further understanding of loss mechanisms and emittance growth can also be gained from particle tracking simulation studies of the full space charge and IBS interplay, followed by benchmarking measurements. Furthermore, dedicated studies of limitations in the LHC are needed, in particular in terms of beam loss handling and collimation. The feasibility of discussed hardware upgrades and operational schemes has yet to be tested experimentally for each ion, in particular for new promising charge states and isotopes. 

From an ion production point of view, detailed LHC filling schemes need to be investigated for cases where the present splitting in the PS is omitted. The efficiency and beam-degrading effects of a possible future LEIR-PS stripper foil call for further studies, as they may introduce unacceptable energy spread or emittance growth. This version of the Injector Model focuses on bunch intensity, but emittance remains a crucial quality factor for luminosity production and should be studied further experimentally across the ion injector chain. Nonetheless, we hope that these findings may bring insight into effective performance-limiting factors in the CERN ion injector chain and in the LHC in order to target ideal ion species for future experiments. 

\section{Acknowledgments}
We thank N. Biancacci, F. Boattini, C. Carli, V. Di Capua, H. Damerau, P. Kruyt, Q. King, D. K\"{u}chler, F. Kr\"{o}ger, J. Olsen, R. Scrivens, M. Seidel, M. Slupecki, T. Stoehlker, and G. Weber for helpful discussions and input during these studies. We acknowledge the support of the Future Ions Working Group at CERN in the execution of studies and allocation of resources.





\bibliographystyle{elsarticle-num2}
\bibliography{ndc_mainbib}

@Article{alice,
  Title                    = {ALICE: Physics Performance Report, Volume I},
  Author                   = {{ALICE Collaboration: F. Carminati} and P Foka and P Giubellino and A Morsch and G Paic and J-P Revol and K Safarik and Y Schutz and {U. A. Wiedemann (editors)}},
  Journal                  = {Journal of Physics G: Nuclear and Particle Physics},
  Year                     = {2004},
  Number                   = {11},
  Pages                    = {1517-1763},
  Volume                   = {30}
}

@Article{atlas,
  Title                    = {The {ATLAS} Experiment at the {CERN Large Hadron Collider}},
  Author                   = {{ATLAS Collaboration}},
  Journal                  = {JINST},
  Year                     = {2008},
  Pages                    = {S08003},
  Volume                   = {3}
}

@Article{lhcdesignV1,
  Title                    = {{LHC} design report v.1 : The {LHC} main ring},
  Author                   = {O. S. Br{\"{u}}ning and P.~Collier and P.~Lebrun and S.~Myers and R.~Ostojic and J.~Poole and {P.~Proudlock (editors)}},
  Journal                  = {CERN-2004-003-V1},
  Address                  = {{CERN, Geneva, Switzerland}},
  Year                     = {CERN, Geneva, Switzerland, 2004}
}

@Article{bruce14_PRSTAB_sixtr,
  Title                    = {{Simulations and measurements of beam loss patterns at the CERN Large Hadron Collider}},
  Author                   = {Bruce, R. and Assmann, R. W. and Boccone, V. and Bracco, C. and Brugger, M. and Cauchi, M. and Cerutti, F. and Deboy, D. and Ferrari, A. and Lari, L. and Marsili, A. and Mereghetti, A. and Mirarchi, D. and Quaranta, E. and Redaelli, S. and Robert-Demolaize, G. and Rossi, A. and Salvachua, B. and Skordis, E. and Tambasco, C. and Valentino, G. and Weiler, T. and Vlachoudis, V. and Wollmann, D.},
  Journal                  = {Phys. Rev. ST Accel. Beams},
  Year                     = {2014},

  Month                    = {08},
  Pages                    = {081004},
  Volume                   = {17},

  Doi                      = {10.1103/PhysRevSTAB.17.081004},
  Issue                    = {8},
  Publisher                = {American Physical Society},
  Url                      = {http://link.aps.org/doi/10.1103/PhysRevSTAB.17.081004}
}

@Article{bruce15_PRSTAB_betaStar,
  Title                    = {{Calculations of safe collimator settings and ${\ensuremath{\beta}}^{*}$ at the CERN Large Hadron Collider}},
  Author                   = {Bruce, R. and Assmann, R. W. and Redaelli, S.},
  Journal                  = {Phys. Rev. ST Accel. Beams},
  Year                     = {2015},

  Month                    = {Jun},
  Pages                    = {061001},
  Volume                   = {18},

  Doi                      = {10.1103/PhysRevSTAB.18.061001},
  Issue                    = {6},
  Numpages                 = {16},
  Publisher                = {American Physical Society},
  Url                      = {http://link.aps.org/doi/10.1103/PhysRevSTAB.18.061001}
}

@Article{prstabBFPP09,
  Title                    = {{Beam losses from ultraperipheral nuclear collisions between Pb ions in the Large Hadron Collider and their alleviation}},
  Author                   = {Bruce, R. and Bocian, D. and Gilardoni, S. and Jowett, J. M.},
  Journal                  = {Phys. Rev. ST Accel. Beams},
  Year                     = {2009},

  Month                    = {Jul},
  Number                   = {7},
  Pages                    = {071002},
  Volume                   = {12},

  Doi                      = {10.1103/PhysRevSTAB.12.071002},
  Numpages                 = {17},
  Publisher                = {American Physical Society}
}

@Article{bruce17_NIM_beta40cm,
  Title                    = {Reaching record-low $\beta^*$ at the {CERN Large Hadron Collider} using a novel scheme of collimator settings and optics },
  Author                   = {R. Bruce and C. Bracco and R. De Maria and M. Giovannozzi and A. Mereghetti and D. Mirarchi and S. Redaelli and E. Quaranta and B. Salvachua},
  Journal                  = {Nucl. Instrum. Methods Phys. Res. A},
  Year                     = {2017},

  Month                    = {Jan},
  Pages                    = {19 - 30},
  Volume                   = {848},

  Doi                      = {http://dx.doi.org/10.1016/j.nima.2016.12.039},
  Url                      = {http://www.sciencedirect.com/science/article/pii/S0168900216313092}
}

@Article{bruce10prstabCTE,
  author    = {Bruce, R. and Jowett, J. M. and Blaskiewicz, M. and Fischer, W.},
  title     = {{Time evolution of the luminosity of colliding heavy-ion beams in BNL Relativistic Heavy Ion Collider and CERN Large Hadron Collider}},
  journal   = {Phys. Rev. ST Accel. Beams},
  year      = {2010},
  volume    = {13},
  pages     = {091001},
  month     = {Sep},
  doi       = {10.1103/PhysRevSTAB.13.091001},
  issue     = {9},
  numpages  = {16},
  publisher = {American Physical Society},
}

@Article{prl07,
  Title                    = {Observations of Beam Losses Due to Bound-Free Pair Production in a Heavy-Ion Collider},
  Author                   = {R. Bruce and J. M.~Jowett and S.~Gilardoni and A.~Drees and W.~Fischer and S.~Tepikian and S. R.~Klein},
  Journal                  = {Phys. Rev. Letters},
  Year                     = {2007},
  Number                   = {14},
  Pages                    = {144801},
  Volume                   = {99},

  Doi                      = {10.1103/PhysRevLett.99.144801},
  Url                      = {https://link.aps.org/doi/10.1103/PhysRevLett.99.144801}
}

@Article{YR_WG5_2018,
  author        = {Citron, Z. and others},
  title         = {{Future physics opportunities for high-density QCD at the LHC with heavy-ion and proton beams}},
  journal       = {{HL/HE-LHC Workshop: Workshop on the Physics of HL-LHC, and Perspectives at HE-LHC Geneva, Switzerland, June 18-20, 2018, CERN-LPCC-2018-07}},
  year          = {2018},
  archiveprefix = {arXiv},
  booktitle     = {{HL/HE-LHC Workshop: Workshop on the Physics of HL-LHC, and Perspectives at HE-LHC Geneva, Switzerland, June 18-20, 2018}},
  comment       = {R. Bruce},
  eprint        = {1812.06772},
  primaryclass  = {hep-ph},
  slaccitation  = {%%CITATION = ARXIV:1812.06772;%%},
  url           = {https://cds.cern.ch/record/2650176?ln=en},
}

@Article{cms,
  Title                    = {The {CMS} Experiment at the {CERN LHC}},
  Author                   = {{CMS Collaboration}},
  Journal                  = {JINST},
  Year                     = {2008},
  Pages                    = {S08004},
  Volume                   = {3}
}

@TechReport{liu_TDR_ions,
  Title                    = {{LHC Injectors Upgrade, Technical Design Report, Vol. II: Ions}},
  Author                   = {Coupard, J. and Damerau, H. and Funken, A. and
 Garoby, R. and Gilardoni, S. and Goddard, B.
 and Hanke, K. and Manglunki, D. and Meddahi, M. 
 and Rumolo, G. and Scrivens, R. and
 Chapochnikova, E.},
  Institution              = {CERN},
  Year                     = {2016},

  Address                  = {Geneva},
  Month                    = {Apr},
  Number                   = {CERN-ACC-2016-0041},

  Reportnumber             = {CERN-ACC-2016-0041},
  Url                      = {https://cds.cern.ch/record/2153863}
}

@Article{fuster18_ipac_Xe,
  Title                    = {{Cleaning performance of the collimation system with Xe beams at the Large Hadron Collider}},
  Author                   = {N. Fuster-Martinez and J. M. Jowett and P. Hermes and R. Bruce and D. Mirarchi and S. Redaelli},
  Journal                  = {Proceedings of the International Particle Accelerator Conference 2018, Vancouver, Canada},
  Year                     = {2018},
  Pages                    = {176},

  Url                      = {http://accelconf.web.cern.ch/AccelConf/ipac2018/papers/mopmf038.pdf}
}

@Article{bartosik19_evian,
  Title                    = {{Injectors beam performance evolution during Run~2}},
  Author                   = {{H. Bartosik \textit{et al.}}},
  Journal                  = {Proceedings of the 9th LHC Operations Evian Workshop, Evian, France},
  Year                     = {2019},

  Url                      = {https://indico.cern.ch/event/751857/timetable/#20190130.detailed}
}

@Article{hermes16_nim,
  Title                    = {Measured and simulated heavy-ion beam loss patterns at the {CERN Large Hadron Collider}},
  Author                   = {P.D. Hermes and R. Bruce and J. M. Jowett and S. Redaelli and B. Salvachua Ferrando and G. Valentino and D. Wollmann},
  Journal                  = {Nucl. Instrum. Methods Phys. Res. A},
  Year                     = {2016},

  Month                    = {Feb},
  Pages                    = {73 - 83},
  Volume                   = {819},

  Doi                      = {http://dx.doi.org/10.1016/j.nima.2016.02.050},
  Url                      = {https://www.sciencedirect.com/science/article/pii/S0168900216002175?via%3Dihub}
}

@Article{jowett19_evian,
  Title                    = {{Overview of ion runs during Run 2}},
  Author                   = {{J. M. Jowett \textit{et al.}}},
  Journal                  = {Proceedings of the 9th LHC Operations Evian Workshop, Evian, France},
  Year                     = {2019},

  Url                      = {https://indico.cern.ch/event/751857/timetable/#20190130.detailed}
}

@Article{jowett16_ipac,
  Title                    = {{The 2015 Heavy-Ion Run of the LHC}},
  Author                   = {{ Jowett} and R. Alemany-Fernandez and R. Bruce and M. Giovannozzi and P.D. Hermes and W. H\"{o}fle and M. Lamont and T. Mertens and S. Redaelli and M. Schaumann and J.A. Uythoven and J. Wenninger},
  Journal                  = {Proceedings of the International Particle Accelerator Conference 2016, Busan, Korea},
  Year                     = {2016},
  Pages                    = {1493},

  Url                      = {http://accelconf.web.cern.ch/AccelConf/ipac2016/papers/tupmw027.pdf}
}

@Article{schaumann18_ipac,
  Title                    = {{First Xenon-Xenon Collisions in the LHC}},
  Author                   = {M. Schaumann and R. Alemany-Fernandez and P. Baudrenghien and T. Bohl and C. Bracco and R. Bruce and N. Fuster-Martinez and M. A. Jebramcik and J. M. Jowett and T. Mertens and D. Mirarchi and S. Redaelli and B. Salvachua and M. Solfaroli and H. Timko and J. Wenninger},
  Journal                  = {Proceedings of the International Particle Accelerator Conference 2018, Vancouver, Canada},
  Year                     = {2018},
  Pages                    = {180},

  Url                      = {http://accelconf.web.cern.ch/AccelConf/ipac2018/papers/mopmf039.pdf}
}

@Article{jowett19_ipac,
  author  = {{J. M. Jowett}   and C. Bahamonde Castro and W. Bartmann and C. Bracco and R. Bruce and J. Dilly and S. Fartoukh and A. Garcia-Tabares and M. Hofer and M.A. Jebramcik and J. Keintzel and A. Lechner and E.H. Maclean and L. Malina and T. Medvedeva and D. Mirarchi and T. Persson and B. Petersen and S. Redaelli and M. Schaumann and M. Solfaroli and R. Tomas and J. Wenninger and J. Coello and E. Fol and N. Fuster-Martinez and E.B. Holzer and A. Mereghetti and B. Salvachua and C. Schwick and M. Spitznagel and H. Timko and A. Wegscheider and D. Wollmann},
  title   = {{The 2018 Heavy-Ion Run of the LHC}},
  journal = {Proceedings of the 10th International Particle Accelerator Conference (IPAC2019): Melbourne, Australia, May 19-24, 2019},
  year    = {2019},
  pages   = {2258},
  doi     = {https://doi.org/10.18429/JACoW-IPAC2019-WEYYPLM2},
}

@Article{ipac11_jowett_fist_Pb_run,
  author  = {{J. M. Jowett} and G. Arduini and R. Assmann and P. Baudrenghien and C. Carli and M. Lamont and M. Solfaroli Camillocci and J. Uythoven and W. Venturini and J. Wenninger},
  title   = {{First run of the LHC as a heavy-ion collider}},
  journal = {Proceedings of IPAC11, San Sebastian, Spain},
  year    = {2011},
  pages   = {1837},
  url     = {http://accelconf.web.cern.ch/AccelConf/IPAC2011/papers/tupz016.pdf},
}

@Article{jowett17_cham,
  author  = {{J. M. Jowett} and R. {Alemany Fernandez} and Michaela Schaumann and P. Hermes and T. Mertens},
  title   = {{HL-LHC beam parameters and performance with heavy ions---an update}},
  journal = {Presentation in the LHC Performance Workshop (Chamonix 2017), Chamonix, France},
  year    = {2017},
  url     = {https://indico.cern.ch/event/580313/timetable/},
}

@techreport{lopienska2022_cern_complex,
  title={The CERN accelerator complex, layout in 2022},
  author={Lopienska, Ewa},
  year={2022},
  note={Available at: \url{https://cds.cern.ch/record/2800984/files/CCC-v2022.png}}
}

@misc{alice_2022_letter_of_intent,
	title = {Letter of intent for {ALICE} 3: {A} next-generation heavy-ion experiment at the {LHC}},
	shorttitle = {Letter of intent for {ALICE} 3},
	url = {http://arxiv.org/abs/2211.02491},
	urldate = {2023-03-03},
	publisher = {arXiv},
	author = {{ALICE Collaboration}},
	month = nov,
	year = {2022},
	note = {arXiv:2211.02491 [hep-ex, physics:nucl-ex, physics:physics]},
}

@InProceedings{NA61_light_ions,
  title={Development of fragmented low-{Z} ion beams for the {NA61} experiment at the {CERN SPS}},
  author={Efthymiopoulos, I and and others},
  booktitle    = {Proceedings of the 2nd International Particle Accelerator Conference (IPAC2011), San Sebastián, Spain},
paper        = {THPS051},
  publisher={{JACoW} Publishing, Geneva, Switzerland},
  year={2011},
}

@article{krasny2015gamma,
  title={The gamma factory proposal for {CERN}},
  author={Krasny, Mieczyslaw Witold},
  journal={arXiv preprint arXiv:1511.07794},
  year={2015}
}

@misc{injector_model_github_repo,
    title = {Injector Model Github repository},
    author = {Elias Waagaard},
    note = {Available at: \url{https://github.com/ewaagaard/injector_model}}
}

@article{baron,
  title={Charge exchange of very heavy ions in carbon foils and in the residual gas of GANIL cyclotrons},
  author={Baron, E and Bajard, M and Ricaud, Ch},
  journal={Nuclear Instruments and Methods in Physics Research Section A: Accelerators, Spectrometers, Detectors and Associated Equipment},
  volume={328},
  number={1-2},
  pages={177--182},
  year={1993},
  publisher={Elsevier}
}

@misc{schindl_space_1999,
	title = {Space charge - introduction},
	url = {https://cds.cern.ch/record/384395},
	urldate = {2022-08-24},
	journal = {CERN Document Server},
	author = {Schindl, Karlheinz},
	month = mar,
	year = {1999},
	doi = {10.5170/CERN-2005-004.285},
	note = {Number: CERN-PS-99-012-DI, Publisher: CERN. Available at \url{https://cds.cern.ch/record/384395}},
}

@Article{bruce20_HL_ion_report,
  author  = {R. Bruce and T. Argyropoulos and H. Bartosik and R. De Maria and N. Fuster-Martinez and M.A. Jebramcik and {J. M. Jowett} and N. Mounet and S. Redaelli and G. Rumolo and M. Schaumann and H. Timko},
  title   = {{HL-LHC} operational scenario for {Pb-Pb} and {p-Pb} operation},
  journal = {CERN-ACC-2020-0011},
  year    = {2020},
  url     = {https://cds.cern.ch/record/2722753},
}

@Article{schaumann20_PRAB_BFPP,
  author    = {Schaumann, M. and Jowett, J. M. and Bahamonde Castro, C. and Bruce, R. and Lechner, A. and Mertens, T.},
  title     = {{Bound-free pair production from nuclear collisions and the steady-state quench limit of the main dipole magnets of the CERN Large Hadron Collider}},
  journal   = {Phys. Rev. Accel. Beams},
  year      = {2020},
  volume    = {23},
  pages     = {121003},
  month     = {Dec},
  doi       = {10.1103/PhysRevAccelBeams.23.121003},
  issue     = {12},
  numpages  = {15},
  publisher = {American Physical Society},
  url       = {https://link.aps.org/doi/10.1103/PhysRevAccelBeams.23.121003},
}

@Article{arek20_PSI_collimation,
  author    = {Gorzawski, A. and Abramov, A. and Bruce, R. and Fuster-Martinez, N. and Krasny, M. and Molson, J. and Redaelli, S. and Schaumann, M.},
  title     = {Collimation of partially stripped ions in the CERN Large Hadron Collider},
  journal   = {Phys. Rev. Accel. Beams},
  year      = {2020},
  volume    = {23},
  pages     = {101002},
  month     = {Oct},
  doi       = {10.1103/PhysRevAccelBeams.23.101002},
  issue     = {10},
  numpages  = {11},
  publisher = {American Physical Society},
  url       = {https://link.aps.org/doi/10.1103/PhysRevAccelBeams.23.101002},
}

@Article{bruce21_EPJplus_ionLumi,
  author  = {R. Bruce and M.A. Jebramcik and {J. M. Jowett} and T. Mertens and M. Schaumann},
  title   = {Performance and luminosity models for heavy-ion operation at the {CERN Large Hadron Collider}},
  journal = {Eur. Phys. J. Plus},
  year    = {2021},
  volume  = {136},
  pages   = {745},
  comment = {ionsHL},
  doi     = {10.1140/epjp/s13360-021-01685-5},
  url     = {https://doi.org/10.1140/epjp/s13360-021-01685-5},
}

@Article{John2021,
  author        = {John, Isabelle and Bartosik, Hannes},
  title         = {{Space charge and intrabeam scattering effects for Lead-ions and Oxygen-ions in the LHC injector chain}},
  journal       = {CERN-ACC-NOTE-2021-0003},
  year          = {2021},
  month         = {Jan},
  __markedentry = {[roderik:]},
  url           = {http://cds.cern.ch/record/2749453},
}

@Article{Bruce2021a,
  author        = {Bruce, R. and Alemany-Fern\'andez, R. and Bartosik, H. and Jebramcik, M. and Jowett, J. M. and Schaumann, M.},
  title         = {{Studies for an LHC Pilot Run with Oxygen Beams}},
  journal       = {Proceedings of the 12th International Particle Accelerator Conference (IPAC'21): Campinas, Brazil, May 2021},
  year          = {2021},
  pages         = {53--56},
  __markedentry = {[roderik:]},
  doi           = {10.18429/JACoW-IPAC2021-MOPAB005},
  paper         = {MOPAB005},
}

@Book{Alonso2020,
  title         = {{High-Luminosity Large Hadron Collider (HL-LHC): Technical design report}},
  publisher     = {CERN},
  year          = {2020},
  author        = {I. Bejar Alonso and O. Bruning and P. Fessia and M. Lamont and L. Rossi and L. Tavian and {M. Zerlauth (editors)}},
  series        = {CERN Yellow Reports: Monographs, CERN-2020-0010},
  address       = {Geneva},
  __markedentry = {[roderik:]},
  comment       = {R. Bruce},
  doi           = {10.23731/CYRM-2020-0010},
  url           = {http://cds.cern.ch/record/2749422?ln=en},
}

@article{kroger2022_gamma_factory_stripping_charge,
  title={Charge-State Distributions of Highly Charged Lead Ions at Relativistic Collision Energies},
  author={Kr{\"o}ger, Felix M and Weber, G{\"u}nter and Hirlaender, Simon and Alemany-Fernandez, Reyes and Krasny, Mieczyslaw W and St{\"o}hlker, Thomas and Tolstikhina, Inga Yu and Shevelko, Viacheslav P},
  journal={Annalen der Physik},
  volume={534},
  number={3},
  pages={2100245},
  year={2022},
  publisher={Wiley Online Library}
}

@article{scheidenberger1998_heavy_ion_penetration,
  title={Penetration of relativistic heavy ions through matter},
  author={Scheidenberger, C and Geissel, H},
  journal={Nuclear Instruments and Methods in Physics Research Section B: Beam Interactions with Materials and Atoms},
  volume={135},
  number={1-4},
  pages={25--34},
  year={1998},
  publisher={Elsevier}
}

@techreport{north_area_requests,
  title={Addendum to the {NA61/SHINE} Proposal: Request for light ions beams in Run 4},
  author={Mackowiak-Pawlowska, M},
  note={NA61/SHINE Collaboration. CERN report: No. CERN-SPSC-2023-022.},
  year={2023}
}

@misc{detlef_Kr_source,
  title={Preliminary results of the source performance with krypton},
  author={Detlef Küchler},
  year={2023},
  note={{Presentation at CERN} New Ions Working Group Update Symposium on deliverables D1.1 and D3.1. Available at: \url{https://indico.cern.ch/event/1304924/}}
}

@misc{detlef_Mg_source,
  title={Magnesium test at the {GTS-LHC} ion source: preliminary results},
  author={Detlef Küchler},
  year={2024},
}

@misc{Scrivens_gamma_factory_source_status,
  title={The {CERN} heavy ion source - present status},
  author={D. Küchler and R. Scrivens},
  year={2019},
  note={Presentation at the Gamma Factory Meeting. Available at: \url{https://indico.cern.ch/event/802131/}}
}

@misc{Kroeger_gamma_factory_alternative_stripper_foil,
  title={Optimization of the stripper foil material and thickness for the {Gamma Factory}},
  author={F. Kröger and others},
  year={2019},
  note={Presentation at the Gamma Factory Meeting. Available at: \url{https://indico.cern.ch/event/802131/}}
}

@inproceedings{2019_sps_recent_Pb_performance,
  title={Recent beam performance achievements with the {Pb} ion beam in the {SPS} for {LHC} physics runs},
  author={Bartosik, Hannes and Alemany, R and Argyropoulos, Theodoros and Bohl, Thomas and Damerau, Heiko and Kain, Verena and Papotti, Giulia and Rumolo, Giovanni and Hernandez, A Saa and Shaposhnikova, Elena},
  doi={10.18429/JACoW-IPAC2019-MOPGW069},
  booktitle={10th Int. Particle Accelerator Conf.(IPAC’19), Melbourne, Australia},
  year={2019}
}

@inproceedings{leir_ecooler,
  title={{LEIR} Cooler Status},
  author={Tranquille, Gerard},
  booktitle={AIP Conference Proceedings},
  volume={821},
  number={1},
  pages={57--64},
  year={2006},
  organization={American Institute of Physics},
  note={Available at \url{https://pubs.aip.org/aip/acp/article-abstract/821/1/57/946361/LEIR-Cooler-Status}}
}

@article{parkhomchuk2000_model,
  title={New insights in the theory of electron cooling},
  author={Parkhomchuk, VV},
  journal={Nuclear Instruments and Methods in Physics Research Section A: Accelerators, Spectrometers, Detectors and Associated Equipment},
  volume={441},
  number={1-2},
  pages={9--17},
  year={2000},
  publisher={Elsevier}
}

@article{parkhomchuk_overview,
  title={Electron cooling: 35 years of development},
  author={Parkhomchuk, Vasilii V and Skrinski{\u\i}, Aleksandr N},
  journal={Physics-Uspekhi},
  volume={43},
  number={5},
  pages={433},
  year={2000},
  publisher={IOP Publishing}
}

@article{leir_ecooling_1999experimental,
  title={Experimental investigation of electron cooling and stacking of lead ions in a low energy accumulation ring},
  author={Bosser, Jacques and Chanel, M and Hill, C and Lombardi, AM and M{\"o}hl, D and Vretenar, Maurizio and Carli, Christian and Maury, S and Tanke, E and Molinari, G and others},
  journal={Part. Accel.},
  volume={63},
  number={CERN-PS-99-033-DI},
  pages={171--210},
  year={1999},
  note={Available at \url{https://cds.cern.ch/record/388111}}
}

@inproceedings{kruyt_xsuite_vs_betacool,
  title={Advancements and applications of cooling simulation tools: A focus on X-suite},
  author={Kruyt, P and Gamba, D and Franchetti, G},
  booktitle={Proceedings of the COOL 2023 Conference, Montreux, Switzerland},
  year={2023},
  organization={CERN}
}

@article{mahner2007_LEIR_vacuum_system,
  title={The vacuum system of the {Low Energy Ion Ring at CERN}: Requirements, design, and challenges},
  author={Mahner, E},
  journal={Vacuum},
  volume={81},
  number={6},
  pages={727--730},
  year={2007},
  publisher={Elsevier}
}

@inproceedings{slupecki_iefc_2024_leir_performance,
    author = {M. Slupecki},
    title = {Update on ion readiness in the injectors},
    booktitle = {Proceedings of the LHC Injector and Experimental Facilities Committee (IEFC) Meeting 18/10/2024} ,
    year = {2024},
    note = {Available at \url{https://indico.cern.ch/event/1464610/}}
}

@inproceedings{ipp_mg7_test_slupecki,
    author = {M. Slupecki},
    title = {Outcome of the {Mg7+} beam test and future plans},
    booktitle = {Proceedings of the CERN Injectors Performance Panel (IPP) Meeting 21/07/2024} ,
    year = {2024},
    note = {Available at \url{https://indico.cern.ch/event/1420323/}}
}

@misc{xsuite,
      title={Xsuite: an integrated beam physics simulation framework}, 
      author={G. Iadarola and R. De Maria and S. Lopaciuk and A. Abramov and X. Buffat and D. Demetriadou and L. Deniau and P. Hermes and P. Kicsiny and P. Kruyt and A. Latina and L. Mether and K. Paraschou and Sterbini and F. Van Der Veken and P. Belanger and P. Niedermayer and D. Di Croce and T. Pieloni and L. Van Riesen-Haupt},
      year={2023},
      eprint={2310.00317},
      archivePrefix={arXiv},
      primaryClass={physics.acc-ph},
      url={https://arxiv.org/abs/2310.00317}, 
}

@phdthesis{papadopoulou_bunch_2020,
	title = {Bunch characteristics evolution for lepton and hadron rings under the inﬂuence of the intra-beam scattering eﬀect},
	url = {https://cds.cern.ch/record/2745710},
	language = {sv},
	urldate = {2022-10-28},
	school = {University of Crete (GR)},
	author = {Papadopoulou, Parthena Stefania},
	month = nov,
	year = {2020},
	note = {Number: CERN-THESIS-2019-405
Publication Title: CERN Document Server},
}

@misc{piwinski_intra-beam-scattering_1974,
	title = {Intra-beam-scattering},
	url = {https://cds.cern.ch/record/400720},
	language = {sv},
	urldate = {2022-10-31},
	journal = {CERN Document Server},
	author = {Piwinski, A.},
	year = {1974},
	doi = {10.5170/CERN-1992-001.226},
	note = {10.5170/CERN-1992-001.226, Publisher: CERN. Available at \url{https://cds.cern.ch/record/400720}},
}

@article{bjorken_intrabeam_1983,
	title = {Intrabeam Scattering},
	language = {en},
	author = {Bjorken, James D},
        journal = {Particle Accelerators},
        vol = {13},
	year = {1983},
	pages = {115-143},
        number={FERMILAB-PUB-82-47-THY},
}

@article{zenkevich_new_2006,
	series = {Proceedings of the {Workshop} on {High} {Intensity} {Beam} {Dynamics}},
	title = {A new algorithm for the kinetic analysis of intra-beam scattering in storage rings},
	volume = {561},
	issn = {0168-9002},
	url = {https://www.sciencedirect.com/science/article/pii/S0168900206000465},
	doi = {10.1016/j.nima.2006.01.013},
	language = {en},
	number = {2},
	urldate = {2022-10-31},
	journal = {Nuclear Instruments and Methods in Physics Research Section A: Accelerators, Spectrometers, Detectors and Associated Equipment},
	author = {Zenkevich, P. and Boine-Frankenheim, O. and Bolshakov, A.},
	month = jun,
	year = {2006},
	pages = {284--288},
}

@article{nagaitsev_intrabeam_2005,
	title = {Intrabeam scattering formulas for fast numerical evaluation},
	volume = {8},
	url = {https://link.aps.org/doi/10.1103/PhysRevSTAB.8.064403},
	doi = {10.1103/PhysRevSTAB.8.064403},
	number = {6},
	urldate = {2022-10-31},
	journal = {Physical Review Special Topics - Accelerators and Beams},
	author = {Nagaitsev, Sergei},
	month = jun,
	year = {2005},
	note = {Publisher: American Physical Society},
	pages = {064403},
}

@TechReport{LIU_ion_targets,
  author = {G. Rumolo},
  title         = {{LIU ion beam parameters table}},
  institution   = {CERN},
  year          = {2017},
  note           = {Report no. LIU-PM-ES-0004 v.2. Available at \url{https://edms.cern.ch/ui/file/1420286/2/LIU-Ions_beam_parameter_table.pdf}},
}

@inproceedings{felix_ibs_study_2024,
    author = {Soubelet, F and others},
    title = {Development of numerical tools for intra-beam scattering modelling},
    booktitle = {Proceedings of the 15th International Particle Accelerator Conf. (IPAC’24), Nashville, United States of America paper 929.},
    year = {2024}
}

@article{hirlaender_xe_lifetime_2018,
	title = {Lifetime and {Beam} {Losses} {Studies} of {Partially} {Strip} {Ions} in the {SPS} ({$^{129}$Xe}$^{39+}$)},
	volume = {IPAC2018},
	copyright = {CC 3.0},
	url = {http://jacow.org/ipac2018/doi/JACoW-IPAC2018-THPMF015.html},
	doi = {10.18429/JACOW-IPAC2018-THPMF015},
	language = {en},
	urldate = {2022-11-11},
	journal = {Proceedings of the 9th Int. Particle Accelerator Conf.},
	author = {Hirlaender, Simon and others},
	collaborator = {Todd (Ed.), Satogata and RW (Ed.), Volker, Schaa},
	year = {2018},
	note = {Artwork Size: 3 pages, 3.305 MB
ISBN: 9783954501847
Publisher: JACoW Publishing, Geneva, Switzerland},
	pages = {3 pages, 3.305 MB},
}

@book{basic_atomic_interactions_book_tolstikhina2018,
  title={Basic Atomic Interactions of Accelerated Heavy Ions in Matter},
  author={Tolstikhina, Inga and Imai, Makoto and Winckler, Nicolas and Shevelko, Viacheslav},
  journal={Springer Series on Atomic, Optical, and Plasma Physics, Springer},
  year={2018},
  publisher={Springer}
}

@article{schlachter_electron_1983,
	title = {Electron capture for fast highly charged ions in gas targets: {An} empirical scaling rule},
	volume = {27},
	issn = {0556-2791},
	shorttitle = {Electron capture for fast highly charged ions in gas targets},
	url = {https://link.aps.org/doi/10.1103/PhysRevA.27.3372},
	doi = {10.1103/PhysRevA.27.3372},
	language = {en},
	number = {6},
	urldate = {2023-04-05},
	journal = {Physical Review A},
	author = {Schlachter, A. S. and Stearns, J. W. and Graham, W. G. and Berkner, K. H. and Pyle, R. V. and Tanis, J. A.},
	month = jun,
	year = {1983},
	pages = {3372--3374},
}

@misc{beam_gas_collisions_github_repo,
    author = {E. Waagaard},
    title = {Beam-Gas Collisions Github repository},
    note = {Available at: \url{https://github.com/ewaagaard/beam_gas_collisions}}
}

@article{shevelko2018lifetimes_ricode_m,
  title={Lifetimes of relativistic heavy-ion beams in the High Energy Storage Ring of {FAIR}},
  author={Shevelko, VP and Litvinov, Yu A and St{\"o}hlker, Th and Tolstikhina, I Yu},
  journal={Nuclear Instruments and Methods in Physics Research Section B: Beam Interactions with Materials and Atoms},
  volume={421},
  pages={45--49},
  year={2018},
  publisher={Elsevier}
}

@misc{weber_2016_semi_empirical_formula,
	title = {A new semi-empirical formular for total electron loss by energetic ions},
	author = {Weber, G.},
	year = {2017},
	note = {2016 Annual Report Helmholtz-Institut Jena. DOI: 10.15120/GSI-2017-00708. Available at \url{http://repository.gsi.de/record/201624}.},
}

@misc{olsen_baroncs_github_repo,
    author = {J. Olsen},
    title = {baroncs Github repository},
    note = {Available at: \url{https://github.com/jako4295/baroncs}}
}

@misc{bruce_2021_Preliminary_LHC_light_ion_scenarios,
    author = {R. Bruce and others},
    title = {{Preliminary LHC light-ion scenarios  for Run 5 and beyond}},
    note = {Presentation LIU-Ions PS injectors Performance Coordination Meeting (2021). Available at: \url{https://indico.cern.ch/event/1085343/}}
}

@article{waagaard_2023_HB,
  title={Exploring Space Charge and Intra-beam Scattering Effects in the {CERN} Ion Injector Chain},
  author={Waagaard, Elias and Bartosik, Hannes},
  journal={JACoW Proceedings of the ICFA Advanced Beam Dynamics Workshop on High-Intensity and High-Brightness Hadron Beams (68th)},
  volume={2023},
  pages={515--518},
  year={2024},
  note={Available at \url{https://accelconf.web.cern.ch/hb2023/doi/JACoW-HB2023-THBP23.html}}}

@article{kondev2021_nuclear_database,
  title={The {NUBASE2020} evaluation of nuclear physics properties},
  author={Kondev, FG and Wang, Meng and Huang, WJ and Naimi, S and Audi, G},
  journal={Chinese Physics C},
  volume={45},
  number={3},
  pages={030001},
  year={2021},
  publisher={IOP Publishing}
}

@misc{ps_rf_upgrades,
    title = {{PS RF} Cavities for 50 ns ion bunch spacing - user specification for the study phase},
    author = {CERN},
    year = {2025},
    note = {Functional specification CPS-A-ES-0001 v. 2.1 - available at \url{https://edms.cern.ch/document/3233949/2.1}},
}







\newpage

\onecolumn
\section*{Appendix}

Table~\ref{table:all_scenarios} shows the results from all scenarios in Fig.~\ref{fig:all_scenarios} for each ion species, considering limits from the LEIR electron cooler. These results include the number of allowed LEIR injections, the maximal injected intensity in each machine, the space charge limit for LEIR and the SPS, as well as injected bunch intensity into the LHC.

\begin{table*}[htb!]
\begin{adjustbox}{width=\textwidth}
\begin{tabular}{cccc|c|cc|cc} \hline
   & \multicolumn{3}{c}{LEIR}                                                                                                              & \multicolumn{1}{c}{PS}                                                                                                                            & \multicolumn{2}{c}{SPS}                                                                                                                           & \multicolumn{2}{c}{LHC}                                                                                                                \\ \hline
      $q_{\mathrm{LEIR}}$ & \begin{tabular}[c]{@{}c@{}}N.o. LEIR\\ injections\end{tabular} & \begin{tabular}[c]{@{}c@{}}Max intensity\\ inj. (ions/bunch)\end{tabular} & \begin{tabular}[c]{@{}c@{}}Space charge \\ limit (ions/bunch)\end{tabular} & \begin{tabular}[c]{@{}c@{}}Max intensity\\ inj. (ions/bunch)\end{tabular} & \begin{tabular}[c]{@{}c@{}}Max intensity\\ inj.(ions/bunch)\end{tabular} & \begin{tabular}[c]{@{}c@{}}Space charge \\ limit (ions/bunch)\end{tabular} & \begin{tabular}[c]{@{}c@{}}Intensity\\ (ions/bunch)\end{tabular} & \begin{tabular}[c]{@{}c@{}}Intensity\\ (charges/bunch)\end{tabular} \\ \hline \hline
\multicolumn{4}{l}{\textbf{1: Baseline scenario (2 extracted bunches in LEIR, 4 in the PS)}}                                                                                          & \multicolumn{1}{l}{}                                                                                & \multicolumn{1}{l}{}                                                 & \multicolumn{1}{l}{}                                                       & \multicolumn{1}{l}{}                                             & \multicolumn{1}{l}{}                                                \\ \hline

He$^{1+}$  & 1                              & 1.5E+11            & 1.4E+11                    & 5.1E+10          & 1.8E+10           & 1.1E+10                   & 7.5E+09           & 1.5E+10              \\
O$^{4+}$   & 1                              & 2.1E+10            & 3.6E+10                    & 7.4E+09          & 2.5E+09           & 2.8E+09                   & 1.7E+09           & 1.3E+10              \\
Mg$^{7+}$  & 1                              & 2.7E+09            & 1.8E+10                    & 9.5E+08          & 3.6E+08           & 3.0E+09                   & 2.4E+08           & 2.9E+09              \\
Ar$^{11+}$ & 1                              & 5.1E+09            & 1.2E+10                    & 1.8E+09          & 7.4E+08           & 1.9E+09                   & 4.9E+08           & 8.9E+09              \\
Ca$^{17+}$ & 4                              & 5.5E+09            & 5.0E+09                    & 1.8E+09          & 7.6E+08           & 5.5E+09                   & 5.1E+08           & 1.0E+10              \\
Kr$^{22+}$ & 3                              & 5.1E+09            & 6.4E+09                    & 1.8E+09          & 7.6E+08           & 8.0E+08                   & 5.1E+08           & 1.8E+10              \\
In$^{37+}$ & 7                              & 4.4E+09            & 3.0E+09                    & 1.1E+09          & 4.6E+08           & 1.1E+09                   & 3.1E+08           & 1.5E+10              \\
Xe$^{39+}$ & 6                              & 4.3E+09            & 3.1E+09                    & 1.1E+09          & 4.6E+08           & 8.8E+08                   & 3.1E+08           & 1.7E+10              \\
Pb$^{54+}$ & 8                              & 4.2E+09            & 2.6E+09                    & 9.1E+08          & 3.9E+08           & 3.9E+08                   & 2.6E+08           & 2.1E+10    \\ \hline   
\multicolumn{4}{l}{\textbf{2: No PS splitting}}                                                                                            & \multicolumn{1}{l}{}                                                 & \multicolumn{1}{l}{}                                                       & \multicolumn{1}{l}{}                                                                                   & \multicolumn{1}{l}{}                                             & \multicolumn{1}{l}{}                                                \\ \hline

He$^{1+}$  & 1                              & 1.5E+11            & 1.4E+11                    & 5.1E+10          & 3.6E+10           & 1.1E+10                   & 7.5E+09           & 1.5E+10              \\
O$^{4+}$   & 1                              & 2.1E+10            & 3.6E+10                    & 7.4E+09          & 5.0E+09           & 2.8E+09                   & 1.9E+09           & 1.5E+10              \\
Mg$^{7+}$  & 1                              & 2.7E+09            & 1.8E+10                    & 9.5E+08          & 7.2E+08           & 3.0E+09                   & 4.8E+08           & 5.8E+09              \\
Ar$^{11+}$ & 1                              & 5.1E+09            & 1.2E+10                    & 1.8E+09          & 1.5E+09           & 1.9E+09                   & 9.9E+08           & 1.8E+10              \\
Ca$^{17+}$ & 4                              & 5.5E+09            & 5.0E+09                    & 1.8E+09          & 1.5E+09           & 5.5E+09                   & 1.0E+09           & 2.0E+10              \\
Kr$^{22+}$ & 3                              & 5.1E+09            & 6.4E+09                    & 1.8E+09          & 1.5E+09           & 8.0E+08                   & 5.4E+08           & 1.9E+10              \\
In$^{37+}$ & 7                              & 4.4E+09            & 3.0E+09                    & 1.1E+09          & 9.2E+08           & 1.1E+09                   & 6.2E+08           & 3.0E+10              \\
Xe$^{39+}$ & 6                              & 4.3E+09            & 3.1E+09                    & 1.1E+09          & 9.3E+08           & 8.8E+08                   & 5.9E+08           & 3.2E+10              \\
Pb$^{54+}$ & 8                              & 4.2E+09            & 2.6E+09                    & 9.1E+08          & 7.8E+08           & 3.9E+08                   & 2.6E+08           & 2.1E+10  \\ \hline 

\multicolumn{4}{l}{\textbf{3: LEIR-PS stripping}}                                                                                          & \multicolumn{1}{l}{}                                                                       & \multicolumn{1}{l}{}                                                 & \multicolumn{1}{l}{}                                                       & \multicolumn{1}{l}{}                                             & \multicolumn{1}{l}{}                                                \\ \hline

He$^{1+}$  & 1                              & 1.5E+11            & 1.4E+11                    & 5.0E+10          & 2.3E+10           & 8.9E+10                   & 1.5E+10           & 3.1E+10              \\
O$^{4+}$   & 1                              & 2.1E+10            & 3.6E+10                    & 7.2E+09          & 3.3E+09           & 2.2E+10                   & 2.2E+09           & 1.8E+10              \\
Mg$^{7+}$  & 1                              & 2.7E+09            & 1.8E+10                    & 9.1E+08          & 4.2E+08           & 1.5E+10                   & 2.8E+08           & 3.4E+09              \\
Ar$^{11+}$ & 1                              & 5.1E+09            & 1.2E+10                    & 1.7E+09          & 8.0E+08           & 8.1E+09                   & 5.4E+08           & 9.7E+09              \\
Ca$^{17+}$ & 4                              & 5.5E+09            & 5.0E+09                    & 1.7E+09          & 7.9E+08           & 9.0E+09                   & 5.3E+08           & 1.1E+10              \\
Kr$^{22+}$ & 3                              & 5.1E+09            & 6.4E+09                    & 1.7E+09          & 7.8E+08           & 3.5E+09                   & 5.2E+08           & 1.9E+10              \\
In$^{37+}$ & 7                              & 4.4E+09            & 3.0E+09                    & 8.7E+08          & 4.0E+08           & 2.7E+09                   & 2.7E+08           & 1.3E+10              \\
Xe$^{39+}$ & 6                              & 4.3E+09            & 3.1E+09                    & 5.8E+08          & 2.7E+08           & 2.3E+09                   & 1.8E+08           & 9.7E+09              \\
Pb$^{54+}$ & 8                              & 4.2E+09            & 2.6E+09                    & 4.9E+08          & 2.3E+08           & 1.4E+09                   & 1.5E+08           & 1.2E+10         \\ \hline 

\multicolumn{5}{l}{\textbf{4: LEIR-PS stripping and no PS splitting}}                                                                                                                                             & \multicolumn{1}{l}{}                                            & \multicolumn{1}{l}{}                                                       & \multicolumn{1}{l}{}                                             & \multicolumn{1}{l}{}                                                \\ \hline

He$^{1+}$  & 1                              & 1.5E+11            & 1.4E+11                    & 5.0E+10          & 4.6E+10           & 8.9E+10                   & 3.1E+10           & 6.1E+10              \\
O$^{4+}$   & 1                              & 2.1E+10            & 3.6E+10                    & 7.2E+09          & 6.6E+09           & 2.2E+10                   & 4.4E+09           & 3.6E+10              \\
Mg$^{7+}$  & 1                              & 2.7E+09            & 1.8E+10                    & 9.1E+08          & 8.4E+08           & 1.5E+10                   & 5.6E+08           & 6.8E+09              \\
Ar$^{11+}$ & 1                              & 5.1E+09            & 1.2E+10                    & 1.7E+09          & 1.6E+09           & 8.1E+09                   & 1.1E+09           & 1.9E+10              \\
Ca$^{17+}$ & 4                              & 5.5E+09            & 5.0E+09                    & 1.7E+09          & 1.6E+09           & 9.0E+09                   & 1.1E+09           & 2.1E+10              \\
Kr$^{22+}$ & 3                              & 5.1E+09            & 6.4E+09                    & 1.7E+09          & 1.6E+09           & 3.5E+09                   & 1.0E+09           & 3.7E+10              \\
In$^{37+}$ & 7                              & 4.4E+09            & 3.0E+09                    & 8.7E+08          & 8.0E+08           & 2.7E+09                   & 5.3E+08           & 2.6E+10              \\
Xe$^{39+}$ & 6                              & 4.3E+09            & 3.1E+09                    & 5.8E+08          & 5.4E+08           & 2.3E+09                   & 3.6E+08           & 1.9E+10              \\
Pb$^{54+}$ & 8                              & 4.2E+09            & 2.6E+09                    & 4.9E+08          & 4.5E+08           & 1.4E+09                   & 3.0E+08           & 2.5E+10    \\ \hline 

\multicolumn{5}{l}{\textbf{5: Optimistic - best LEIR charge state, no PS splitting}}                                                                                                                                           & \multicolumn{1}{l}{}                                                 & \multicolumn{1}{l}{}                                                       & \multicolumn{1}{l}{}                                             & \multicolumn{1}{l}{}                                                \\ \hline

He$^{1+}$  & 1                              & 1.5E+11            & 1.4E+11                    & 5.1E+10          & 3.6E+10           & 1.1E+10                   & 7.5E+09           & 1.5E+10              \\
O$^{5+}$   & 2                              & 3.4E+10            & 2.3E+10                    & 8.2E+09          & 5.2E+09           & 5.5E+09                   & 3.5E+09           & 2.8E+10              \\
Mg$^{7+}$  & 2                              & 5.3E+09            & 1.8E+10                    & 1.9E+09          & 1.4E+09           & 3.0E+09                   & 9.7E+08           & 1.2E+10              \\
Ar$^{11+}$ & 2                              & 1.0E+10            & 1.2E+10                    & 3.6E+09          & 3.0E+09           & 1.9E+09                   & 1.2E+09           & 2.2E+10              \\
Ca$^{17+}$ & 5                              & 6.9E+09            & 5.0E+09                    & 1.8E+09          & 1.5E+09           & 5.5E+09                   & 1.0E+09           & 2.0E+10              \\
Kr$^{28+}$ & 6                              & 4.7E+09            & 4.0E+09                    & 1.4E+09          & 1.2E+09           & 1.6E+09                   & 8.0E+08           & 2.9E+10              \\
In$^{36+}$ & 7                              & 4.3E+09            & 3.2E+09                    & 1.1E+09          & 9.7E+08           & 1.1E+09                   & 6.5E+08           & 3.2E+10              \\
Xe$^{40+}$ & 8                              & 5.4E+09            & 2.9E+09                    & 1.0E+09          & 8.8E+08           & 9.5E+08                   & 5.9E+08           & 3.2E+10              \\
Pb$^{56+}$ & 8                              & 1.9E+09            & 2.4E+09                    & 6.8E+08          & 5.9E+08           & 4.4E+08                   & 2.9E+08           & 2.4E+10             \\ \hline 

\end{tabular}
\end{adjustbox}
\caption{Calculated intensity and space charge limitations through the injector chain for the 1) baseline scenario, 2) No PS splitting, 3) LEIR-PS stripping,  4) LEIR-PS stripping and no PS splitting and 5) best LEIR charge state (from scan) and no PS splitting.}
\label{table:all_scenarios}
\end{table*}

\newpage

Figure~\ref{fig:LEIR_charge_state_scan} shows the projected bunch intensity $N_b$ injected into LHC for different LEIR charge states.

\begin{center}
\begin{figure}[htb!]
\centering
\includegraphics[width=.93\textwidth]{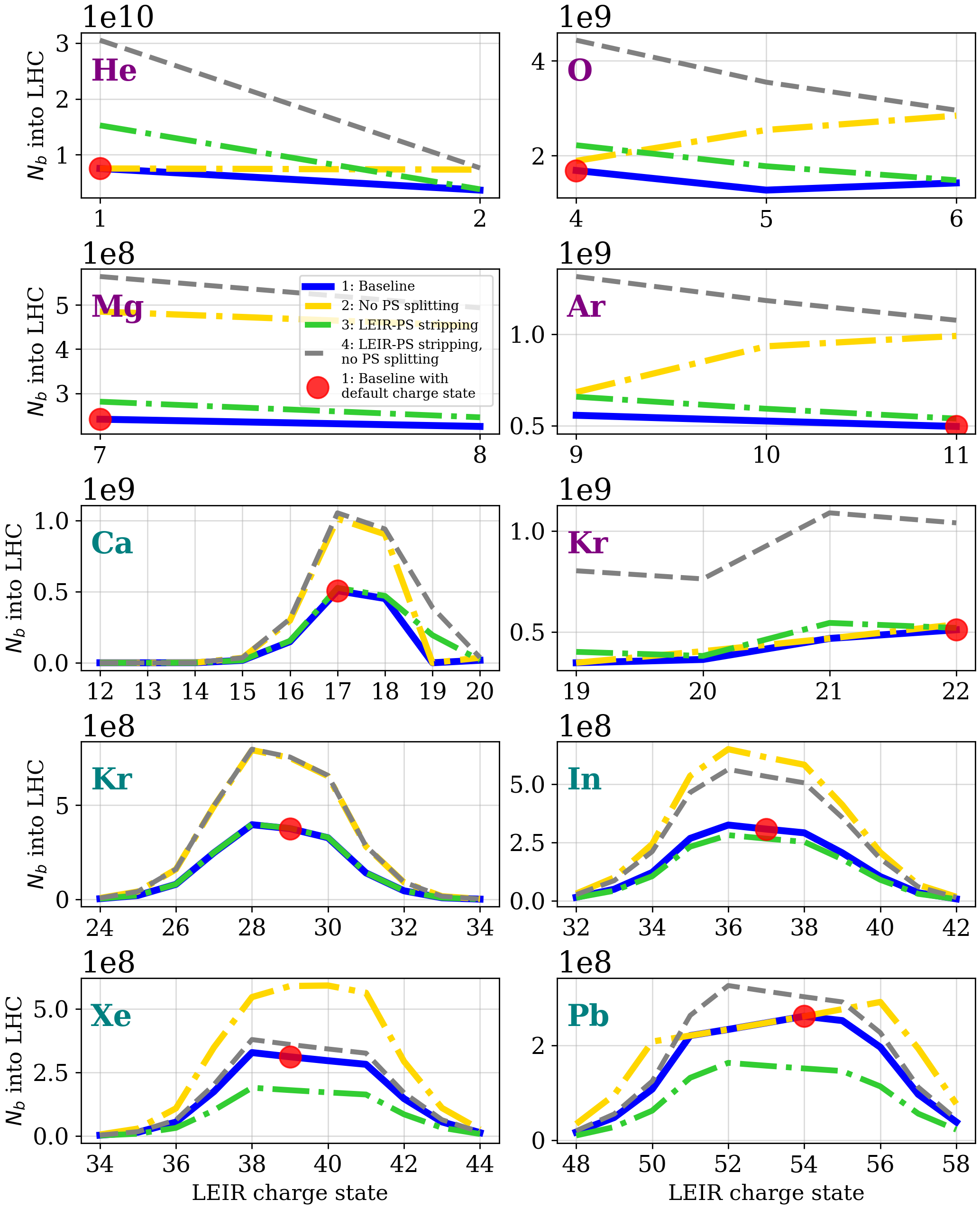}
\caption{LEIR charge state scan for all the considered ions in the first four scenarios. \textcolor{violet}{Purple} ion labels indicate charge state scans directly produced from the source, \textcolor{teal}{teal} ion labels indicate scans charge states produced through stripping after Linac3.}
\label{fig:LEIR_charge_state_scan}
\end{figure}
\end{center}

Figure~\ref{fig:combined_isotope_scan} shows the projected nucleons per bunch $\mathcal{N}$ injected into the LHC scanning over all stable isotopes, assuming that they can be acquired, stably produced by the source and handled by the injectors. 
\begin{center}
\begin{figure}[htb!]
\centering
\includegraphics[width=.87\textwidth]{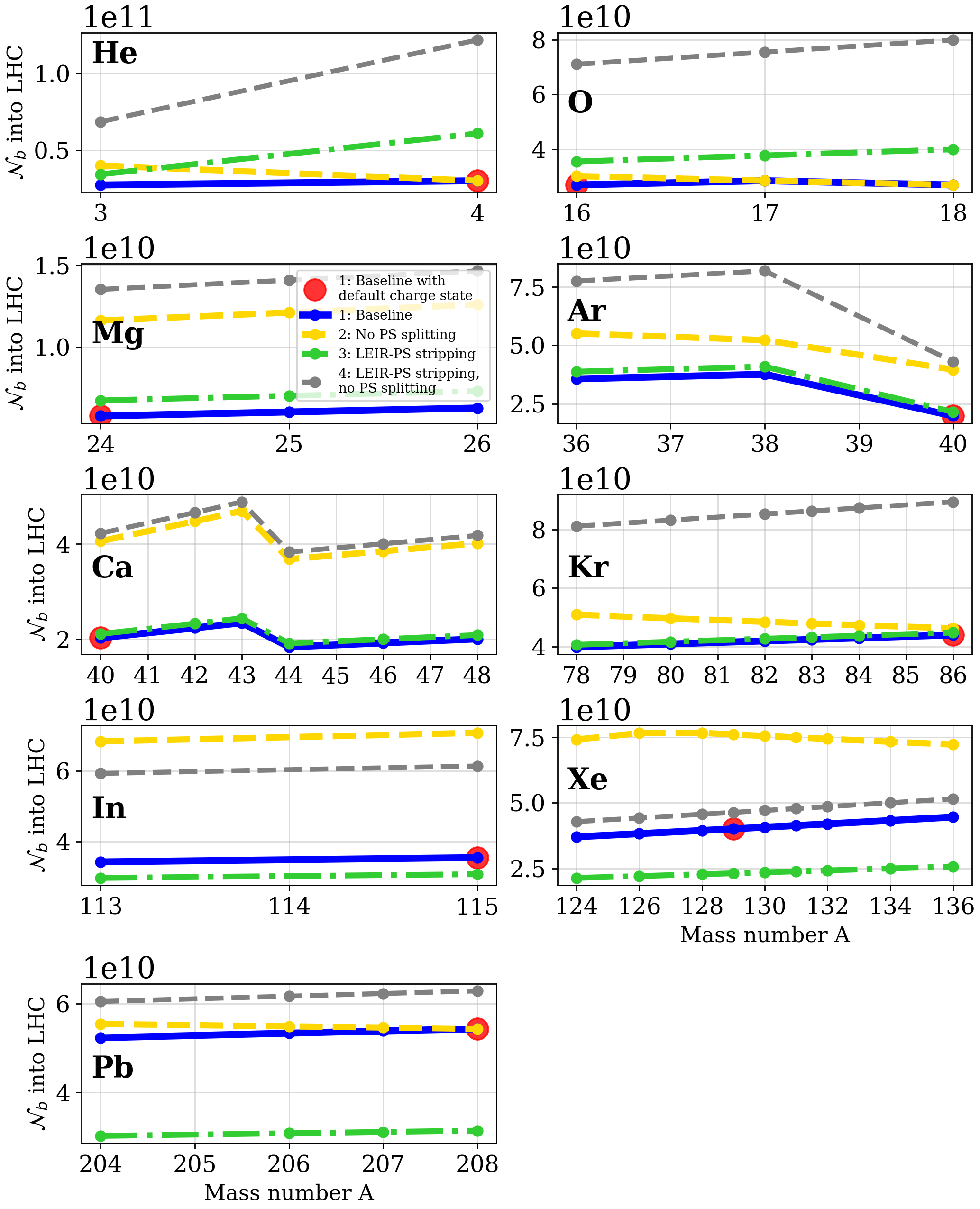}
\caption{Isotope scan for all the stable isotopes in the scenarios considered . The sudden reduction in injected   nucleons per LHC bunch for heavier isotopes of, e.g., Ar and Ca, is explained by fewer allowed LEIR injections due to the longer cooling times required. }
\label{fig:combined_isotope_scan}
\end{figure}
\end{center}

\begin{table*}[ht]
\centering
\caption{Initial intensity and emittance in LHC collisions as obtained from simulations of the beam evolution at the LHC injection plateau. The ion-independent contribution of 97\% to the transmission, as observed in 2024 PbPb run, is included in the intensity calculations.}
\begin{adjustbox}{width=.85\linewidth}
\centering
\begin{tabular}{c|ccc|ccc|ccc} \hline
   & \multicolumn{3}{c}{Geometric emittance [m]} & \multicolumn{3}{c}{Normalised emittance [um]} & \multicolumn{3}{c}{Intensity [ions/bunch]} \\ \hline
   & Baseline            & Optimistic            & 25 ns              & Baseline            & Optimistic            & 25 ns              & Baseline     & Optimistic     & 25 ns                   \\ \hline \hline
O  & 7.9E-10 & 7.7E-10 & 7.9E-10 & 2.9 & 2.9 & 2.9 & 1.62E+09 & 3.3E+09 & 1.6E+09 \\
Ar & 7.9E-10 & 7.8E-10 & 7.9E-10 & 2.6 & 2.7 & 2.6 & 4.7E+08 & 1.1E+09  & 4.6E+08 \\
Ca & 7.9E-10 & 7.8E-10 & 8.0E-10 & 2.9 & 2.9 & 2.9 & 4.8E+08 & 9.2E+08  &  4.8E+08   \\
Kr & 8.9E-10 & 9.0E-10 & 9.0E-10 & 2.7 & 2.8 & 2.8 & 4.8E+08 & 7.0E+08  &  4.6E+08 \\
In & 8.0E-10 & 8.3E-10 & 8.3E-10 & 2.5 & 2.7 & 2.6 & 2.9E+08  & 5.4E+08  &  2.8E+08  \\
Xe & 8.1E-10 & 8.4E-10 & 8.3E-10 & 2.5 & 2.6 & 2.5 & 2.8E+08    & 4.9E+08     & 2.7E+08 \\
Pb & 8.3E-10 & 8.5E-10 & 8.7E-10 & 2.4 & 2.5 & 2.5 &  2.3E+08  & 2.4E+08    & 2.1E+08 \\ \hline
\end{tabular}
\end{adjustbox}
\label{table:collisions_initial_conditions}
\end{table*}

\end{document}